\documentclass[usenatbib]{mn2e}

\usepackage[dvips]{graphicx}
\def\insitu{{\it in situ~}}
\begin{document}

\title[Chemical properties of stellar haloes]
  {Stellar haloes of simulated Milky Way-like galaxies: Chemical and
kinematic properties}

\author[Tissera et al. ]{Patricia B. Tissera $^{1,2}$, Cecilia
  Scannapieco$^{3}$, Timothy C. Beers $^{4}$, Daniela Carollo$^{5}$\\
$^1$  Consejo Nacional de Investigaciones Cient\'{\i}ficas y T\'ecnicas, CONICET, Argentina.\\
$^2$ Instituto de Astronom\'{\i}a y F\'{\i}sica del Espacio, Casilla de Correos 67, Suc. 28, 1428, Buenos Aires, Argentina.\\
$^3$ Leibniz-Institut für Astrophysik Potsdam (AIP), An der Sternwarte 16, D-14482 Potsdam, Germany.\\
$^4$ National Optical Astronomy Observatory, Tucson, Arizona, 85719, USA.\\
$^5$ Macquarie University, Dept. Physics \& Astronomy, Sydney, 2109 NSW, Australia.
}

\maketitle

\begin{abstract}

We investigate the chemical and kinematic properties of the diffuse
stellar haloes of six simulated Milky Way-like galaxies from the
Aquarius Project. Binding energy criteria are adopted to define two
dynamically distinct stellar populations: the diffuse inner and outer
haloes, which comprise different stellar sub-populations with particular
chemical and kinematic characteristics. Our simulated inner- and
outer-halo stellar populations have received contributions from debris
stars (formed in sub-galactic systems while they were
outside the virial radius of the main progenitor galaxies) and endo-debris stars (those
formed in gas-rich sub-galactic systems inside the dark matter haloes
 of the main progenitor galaxy). The inner
haloes possess an additional contribution from disc-heated stars, in the
range $\sim 3 - 30 \% $, with a mean of $\sim 20\% $. Disc-heated stars
might exhibit signatures of kinematical support, in particular among the
youngest ones. Endo-debris plus disc-heated stars define the so-called
\insitu  stellar populations. In both the inner- and outer-halo stellar populations, we detect
contributions from stars with moderate to low [$\alpha$/Fe] ratios,
mainly associated with the endo-debris or disc-heated sub-populations. The
observed abundance gradients in the inner-halo regions are influenced by
both the level of chemical enrichment and the relative contributions
from each stellar sub-population. Steeper abundance gradients in the
inner-halo regions are related to contributions from the disc-heated and
endo-debris stars, which tend to be found at lower binding energies than
debris stars. In the case of the outer-halo regions, although [Fe/H]
gradients are relatively mild, the steeper profiles arise primarily due
to contributions from stars formed in more massive satellites, which
sink farther into the main halo system, and tend to have higher levels
of chemical enrichment and lower energies. Our findings support the
existence of (at least) two distinct diffuse stellar halo populations,
as suggested by a number of recent observations  in the Milky Way and M31. Our results also
indicate that a comparison of the range of predicted kinematics,
abundance gradients, and frequency of [$\alpha$/Fe]-deficient stars with
observations of these quantities in the Milky Way, M31, and other large
spirals can both provide clues to improve the modelling of baryonic
physics, and reveal detailed information about their likely history of
formation and evolution.
\end{abstract}


\section{Introduction}
 
The current cosmological paradigm postulates that large galaxies formed
from the hierarchical aggregation of smaller ones \citep[e.g.,][]{wr78}.
Within this context, a galaxy such as the Milky Way (MW) is expected to
have an assembly path where infall, mergers, and interactions
contribute, with their different relative importance influenced by the
local environment. Cosmological simulations are the most suitable tool
to describe this complex history of formation, and to confront the
adopted cosmological scenario and baryonic sub-grid models with
observations. Sophisticated numerical codes have been developed over the
past few years, and they are continuously being updated to improve their
representation of different physical processes \citep[e.g.,
][]{sprin2005,stin2006,scan09, schaye2011, guedes2011,brook2012}. In
particular, the treatment of chemical evolution within cosmological
codes provides a powerful tool for exploring galaxy formation, and to
confront models and observations \citep[e.g.,][]{mosc2001, lia2002,
gove2007}. In this regard, stellar haloes, such as those of the MW and
M31, are valuable laboratories for confronting expectations from
cosmological models, since most of their stars are quite old, and were
born in the first epochs of galaxy formation. Their chemical abundance
patterns, preserved basically unchanged over their long lifetimes,
provide a fossil record of the physical characteristics of the
interstellar medium at the time and place of their birth \citep[e.g.,
][]{freeman}. We also acknowledge the need to better develop the sets of
questions to be posed to models, so that their full utility for
exploration of the expected behaviour of actual galaxies can be
realized. In this paper, we make a first effort in this direction.

Recent observations provide evidence that the MW halo system comprises
at least two overlapping stellar components, an inner-halo population
and an outer-halo population, with different metallicities, kinematics,
and spatial density profiles \citep[see, e.g., ][]{caro2007,caro2010,
dejong2010,jofre2011,beers2012,kin2012,an2013,hat2013,kaf2013}. These
findings suggest a dual history of halo formation, with contributions
from stars accreted from lower-mass sub-galactic systems and those formed
\insitu. Important additional information on the nature of the halo of
the MW has begun to emerge from consideration of the chemical abundances
of the light and heavy elements. For example, \citet{caro2012} have
shown that clear differences exist between the frequencies of the
so-called carbon-enhanced metal-poor (CEMP) stars that they
kinematically associate with the inner- and outer-halo populations. They
also present evidence for the existence of a correlation between the
frequency of CEMP stars and distance from the Galactic plane, a result
which is difficult to understand in the absence of (at least) a dual
halo. \citet{roederer2009} has argued that kinematically separated
members of the inner- and outer-halo population exhibit a mean [Mg/Fe]
ratio for the outer halo that is lower than that of the inner halo, by
about 0.1 dex. He also demonstrated that the star-to-star scatter in
[Ni/Fe] and [Ba/Fe] for stars associated with the inner halo is
consistently smaller than that of the outer halo. The existence of
different stellar populations located within the inner-halo region of
the MW has recently emerged from studies of the $\alpha$-elements
\citep[e.g., ][]{nissen2010, sheffield2012}. Clearly separable
correlations have been shown to exist between [$\alpha$/Fe] and [Fe/H]
(`high-$\alpha$' and `low-$\alpha$' sequences), as well as with
kinematics. Most recently, \citet{schuster2012} have extended these
differences to include ages. 

State-of-the-art observations are also providing information on the
stellar haloes of nearby galaxies. The nature of the stellar halo of M31
has only recently become feasible to explore in detail, due to
large-scale surveys enabled by modern detectors as well as deep Hubble
Space Telescope observations\citep[e.g.,][]{saraje2012,williams2012}.
Sarajedini et al. presented evidence that the M31 spheroid possesses a
bimodal metallicity distribution, similar in many ways to that suggested
by \citet{caro2007, caro2010} to apply to the MW. In addition, M32 has
been reported to exhibit different spheroid populations, with the outer
one being $~ 1.3$ Gyr older and $\sim 0.2 $ dex more metal poor than its
inner halo \citep{brown2007}.

From a theoretical point of view, cosmological simulations of galaxy
formation within the concordance $\Lambda$-CDM model have also made
progress in the description of the formation history of stellar haloes.
\citet{zolo2009} showed a dual history of formation for stellar haloes,
with contributions from both accreted sub-galactic systems and \insitu star
formation, the latter principally located in the central regions of the
haloes. More recently, similar findings have been reported by
\citet{font2011}, \citet{mcc2012} and \citet{tissera2012}, 
through the analysis of stellar haloes in simulated galaxies using a
variety of numerical codes and sub-grid physics models. Considering
the different approaches used for the baryonic sub-grid models among
these cosmological codes, it is noteworthy that they all agree on
suggesting a dual formation history for the stellar haloes, in the sense
that they are formed by the combination of stars born \insitu and in
accreted sub-galactic systems. This simple way of describing a complex
assembly process, which depends strongly on the galaxy evolutionary
path, has been successful at illustrating the global picture. To move
forward, in this work we aim at studying in more detail the origin of
each different stellar populations that form the diffuse haloes in a
$\Lambda$-CDM scenario.

The results discussed in this paper are related to those reported by
\citet[hereafter, Paper I]{tissera2012} who analysed the chemical
properties of galaxies formed in eight MW mass-size systems of the
Aquarius Project, with different assembly paths within the $\Lambda$-CDM
scenario \citet{scan09}. Paper I shows that, although the central
regions of these simulated galaxies and their stellar haloes are
dominated by very old stars, their chemical properties differ as a
consequence of their individual formation histories. The former were
dominated by \insitu star formation, the latter were assembled mainly
from small accreted sub-galactic systems. Although the mass scales and time
scales involved resulted in different chemical patterns at $z=0$, these
simulations successfully reproduced the global chemo-dynamical trends
observed in the MW.

Following Paper I, we adopt binding energy criteria defined to
dynamically separate two stellar halo populations without any
assumptions on the chemical abundances or kinematics of the stars.
Hereafter, to clarify the description of our results, the most
gravitationally bound stars, dynamically associated with the so-called
inner haloes, will be referred to as inner-halo populations (IHPs), while
the less bound stars, associated with the so-called outer haloes, will
be referred to as outer-halo populations (OHPs). Note that, as the
separation of these stellar components is performed on the basis of
their binding energy content, the IHPs and OHPs can be spatially
superposed. We thus also distinguish these spatial differences by
referring to the inner-halo regions (IHRs) and outer-halo regions
(OHRs); the relative dominance of the different stellar populations
found within these regions naturally differs from one simulated galaxy
to the next, depending on its formation history. The IHR and OHR
  are defined by the spatial distributions of the IHPs and OHPs. The
  crossing radius between the density distributions of these two 
  populations  is $\sim 20$ kpc
  (Tissera et al., in preparation).

To understand their history of assembly, we classify the stars in the
IHPs and OHPs into three sub-populations according to their sites of
formation:

 (1) {\it debris} stars are formed in separate sub-galactic systems
  located outside the virial radius of the main progenitor
   galaxies\footnote{We define the main galaxy as the most massive
     galaxy within the virial radius at $z=0$, and its main progenitor as the most
   massive sub-galactic system from where it formed, as a function of
   redshift.} at the time they were born, and were later accreted onto
   the main galaxy (i.e. stars from dry mergers
as well as the existing populations from wet mergers),

 (2) {\it disc-heated} stars are formed in the disc structure of the
 main progenitor galaxies, then heated up
by some violent event, and 

(3) {\it endo-debris} stars are those formed
in gas-rich sub-galactic systems  located within the virial radius of the
  main progenitor galaxies at their time of formation (i.e. wet mergers).

 In  Paper I, disc-heated and
endo-debris stars were classified as the \insitu component.  In
  this paper, we 
distinguish them as separate sub-populations because, at least in our simulations, they exhibit
different properties which might help link observations with
galaxy formation models.  We note that
  endo-debris stars could be taken either as part of  the \insitu population, because they
  formed within the virial radius of the main progenitor galaxy, or as
  part of the 
  debris population, because they were born from gas associated with
a sub-galactic system. Although we analyse the three sub-populations
separately, when required and to be consistent with Paper I, we adopt the
former interpretation.
Note that we have cleaned our haloes from
bound substructures that could be isolated in density phase. These
substructures will be considered in a separate paper.

This paper is organized as follows. In Section 2, we describe the
simulations and the main aspects of our chemical-enrichment model. This
section also explains the criteria adopted to separate different stellar
components. In Section 3, we study the chemical and kinematic properties
of the different stellar sub-populations contributing to the IHPs, In
Section 4, we discuss the nature of the OHPs. We summarize our findings
in Section 5.

\section{The simulated Milky Way-like systems}

We analyse a suite of six high-resolution simulations of MW galaxy-sized
systems of the original sample from the Aquarius Project,  which include the physics of baryons as described by
\citet{scan09}. We first
selected six out of the eight cases discussed by \citet{scan09}, because two
haloes experienced important mergers in the recent past, unlike in the
MW. The six selected haloes (A, B, C, D, G and H) have surviving discs
of different bulge-to-total mass ratios \citep[see][ and Paper
I]{scan11}.

The target haloes were selected from a cosmological volume consistent
with a $\Lambda$-CDM cosmogony, with parameters $\Omega_{\rm m}=0.25,
\Omega_{\Lambda}=0.75, \sigma_{8}=0.9, n_{s}=1$ and $H_0 = 100 \ h \ { \rm km s^{-1}
Mpc^{-1}}$, and $h =0.73$  \citep[see][]{sprin2008} for further details on
the generation of the initial conditions). The simulations were run with
a version of {\small GADGET-3} \citep{sprin2005}, which includes a
multiphase medium, SN energy feedback, and chemical evolution, as
described in \citet{scan05, scan06}. Maximum gravitational softenings in
the range $\epsilon_{G} =0.5 - 1$ kpc $h^{-1}$ were adopted. Virialized
structures are identified by requiring the mean overdensity to reach
$\approx 200$ times the cosmological critical density. The simulated
haloes have $\approx 1$ million total particles within the virial
radius, and a virial mass in the range $\approx 5 - 11 \times 10^{11}
{\rm M_\odot} h^{-1}$ at $z=0$. Dark matter particles carry masses on
the order of $\approx 10^{6}{\rm M_\odot }h^{-1}$, while, initially, gas
particles possess a total mass of $\approx 2 \times 10^{5}{\rm M_\odot}
h^{-1}$. Their characteristic parameters are summarized in table 1 and
table 2 of \citet[see also Scannapieco et al. 2009]{tissera2010}.

\subsection{The numerical code}

The version of {\small GADGET-3} used to run these simulations includes
a multiphase model for the gas component, metal-dependent cooling, star
formation, and SN feedback, as described in \citet{scan05} and
\citet{scan06}. These multiphase and SN-feedback models have been
used to successfully reproduce the star-formation activity of galaxies
during quiescent and starburst phases, and are able to drive violent
mass-loaded galactic winds with a strength reflecting the depth of the
potential well \citep{scan06, dero10}. This physically-motivated thermal
SN-feedback scheme does not include any ad hoc mass-scale-dependent
parameters, and has no requirement to switch off cooling or provide a
dynamical ``kick'' to other particles. As a consequence, it is
particularly well-suited for the study of galaxy formation in a
cosmological context. 

Our chemical evolution model includes the enrichment by Type~II and
Type~Ia Supernovae (SNII and SNIa, respectively). A Salpeter Initial
Mass Function is assumed, with lower and upper mass cut-offs of 0.1
${\rm M_{\odot}}$ and 40 ${\rm M_{\odot}}$, respectively. We follow 12
different chemical isotopes: H, $^4$He, $^{12}$C, $^{16}$O, $^{24}$Mg,
$^{28}$Si, $^{56}$Fe, $^{14}$N, $^{20}$Ne, $^{32}$S, $^{40}$Ca, and
$^{62}$Zn. Initially, baryons are in the form of gas with primordial
abundance, X$_{\rm H}=0.76$ and Y$_{\rm He}=0.24$.

SNII are considered to originate from stars more massive than 8 ${\rm
M_{\odot}}$. Their nucleosynthesis products are adopted from the
metal-dependent yields of \citet{WW95}. The lifetimes of SNII are
estimated according to the metal-and-mass-dependent lifetime-fitting
formulae of \citet{rait1996}. For SNIa, we adopt the W7 model of
\citet{thie1993}, which assumes that SNIa events originate from CO white
dwarf systems in which mass is transferred from the secondary to the
primary star until the Chandrasekhar mass is exceeded, and an explosion
is triggered. For simplicity, we assume that the lifetime of the
progenitor systems are selected at random over the range $[0.7, 1.1]$
Gyr. To calculate the number of SNIa, we adopt an observationally
motivated relative ratio of SNII to SNIa rates, as explained by 
\citet{mosc2001}.
  
The ejection of chemical elements is grafted onto the SN-feedback model,
so that chemical elements are distributed within the cold and hot gas
phases surrounding a given star particle. The fraction of elements going
into each phase is regulated by a free parameter, $\epsilon_{\rm c}$,
which also determines the amount of SN energy received by each phase.
While the injection of energy follows two different paths, depending on
the thermodynamical properties of the gas, chemical elements are
injected simultaneously with the occurrence of the SN event. All of
these simulations were run with $\epsilon_{\rm c}=0.5$.

\begin{table*}
 \begin{center}
\caption{ Median Abundances of  IHPs and OHPs. First column shows
the encoding name of the simulations. The percentage of stellar
mass, median [Fe/H] and median [O/Fe] for stars identified as
disc-heated, endo-debris and debris sub-populations.  }
\label{tab1}
\begin{tabular}{|l|ccccccccccccccc}\hline

{Systems}   & \multicolumn{9}{c}{ IHPs} & \multicolumn{6}{c}{ OHPs}\\
               & \multicolumn{3}{c}{ Disc-heated}& \multicolumn{3}{c}{   Endo-debris} &\multicolumn{3}{c}{ Debris} & \multicolumn{3}{c}{Endo-debris} &\multicolumn{3}{c}{ Debris}\\                
               & \% &   [Fe/H]&[O/Fe] &  \% &   [Fe/H]&[O/Fe] & \% &   [Fe/H]&[O/Fe] & \% &   [Fe/H]&[O/Fe]& \% &   [Fe/H]&[O/Fe]\\\hline   
Aq-A-5&	31  &  -1.11  &  0.18 &   41 &-1.26  & -0.10   &   25 &-1.45&0.35 & 20 &  -1.94& 0.02& 80 & -1.42 &   0.18  \\
Aq-B-5& 3  & - &- &  29 &  -1.34 &    0.01  &   67 & -1.22   & 0.15 & 12 &    -1.31&  0.02 &88&-1.67&     0.19\\
Aq-C-5& 24 &  -0.78 & -0.05  &32 &-1.37 &   0.01 &   44 &-1.20 &  0.31 &  21 &     -1.90&  0.05 &79&-1.29&     0.12\\
Aq-D-5&	26& -0.90  & -0.06  &  30 & -1.28 &  -0.09 &  43&-1.02 &  0.16 & 24  &     -1.75&   0.02&76 &-1.32&     0.15\\
Aq-G-5 & 35 &-0.90 &  0.00 &  23 & -1.21   & 0.07  &41   &-1.00 &   0.07& 5&  -1.57&  0.02 & 93&-1.81&   0.18\\
Aq-H-5 & 18 & -0.80 &   0.02  & 35 &-0.74 &   -0.14 &45  &-0.86&0.16& 29&     -1.16&   0.02&   71 &-1.17&     0.10 \\
\end{tabular}
   \end{center}
\end{table*}


\subsection{The simulated Aquarius galaxies} 

In order to analyse the chemical properties of the stars, and to relate
them to the assembly history of the final galaxy, Paper I defined
different dynamical components by using the binding energy ($E$) and
angular momentum content along the rotational axis, $J_{z}$. Discs are
defined by those particles with $J_{z}/J_{z, \rm max(E)} > 0.65$ , where
$J_{z, \rm max(E)}$ is the maximum $J_z$ over all particles of a given
binding energy $E$. The rest
of the stellar particles are taken as part of the spheroidal component.
Taking into account their binding energies, stellar particles are
classified as belonging to the central spheroid (`bulge') or the inner
and outer haloes, so that more energetic particles end up in the halo
systems while the less energetic particles define the central spheroids.
Although the classification of stars belonging to the inner and outer
haloes is made on the assumption of a certain binding energy (as
explained in Paper I), variations of this limit only weakly affect our
results, if these variations are sensible.
By applying this methodology, the dynamical decomposition of the stellar populations of the main
  progenitor galaxies into discs, central spheroids, IHPs and OHPs
  was performed from $z\approx 4$ to $z =0$ for all of the Aquarius simulated systems.
 We also cleaned the stellar
haloes of substructure, as defined by the SUBFIND algorithm, so that the
analysed stellar haloes correspond to the diffuse components in density
space. 

As shown in Paper I, the different dynamical components receive
contributions from stars formed both {\it in situ} and in 
sub-galactic systems that were later accreted into the potential well of the
progenitor systems.   As defined in the Introduction,  the \insitu
component is split into two categories: the disc-heated and endo-debris
sub-populations. 
The latter are formed in  gas-rich sub-systems which could
  survive farther into the dark matter haloes and 
continued their star formation activity. New
stars formed in this environment would be more chemically enriched by
SNIa than older ones, particularly if these sub-galactic systems had bursty
star formation histories.
The fractions of each stellar  sub-population (debris, endo-debris
  and disc-heated) associated with each dynamical
component (IHP or OHP) is different, and varies from system-to-system, depending on
their history of assembly. 

Paper I also demonstrated that the OHPs are mainly formed from debris
stars, along with a small fraction of \insitu stars. In order to
understand the origin of these stars, we followed each of them back in
time, and identified the first sub-galactic system that hosted them. We
found that most of the stars that were originally classified as \insitu
stars were in fact formed {\it in between} available snapshots of the
simulations, which complicated their detection. From seeking the
original gas particles in which they formed, it is clear that most of
them were part of small sub-galactic systems. A negligible fraction of
the stars (and their progenitor gas particles) could not be related to
any system, but all of them are old, with formation redshift $z > 5$. At
such high redshifts, our simulations do not possess sufficient
resolution to reliably follow the evolution of the smallest
clumps, so we include them within the endo-debris populations.
Thus, we will hereafter consider that the OHPs are built up mainly by
debris stars from accreted sub-galactic systems,  with  a small
fraction  corresponding to endo-debris stars.

A similar analysis was carried out for the \insitu stars in the IHPs. We
found that these stars included endo-debris sub-populations as well as
disc-heated stars. The latter are stars formed in the disc components,
but that were heated sufficiently to become part of inner haloes by $z
\sim 0 $, contributing from $\sim 3 \%$ to $\sim 30 \%$ of the IHPs.
Disc-heated stars are defined by checking if  they were
  formed from cold gas which was part of  the disc component of the
  main progenitor galaxy at the time they were born.
To do that we traced each star particle back in time, and examined to
which dynamical component it belonged at that time. This was possible because we
defined disc and halo components at every available snapshot of the
simulations, as discussed above.
These fractions are in agreement with the recently reported $\sim 20\%$
of disc-heated stellar fraction estimated for the MW halo by
\citet{sheffield2012}. Endo-debris stars contribute $\approx 30 \%$
of the IHPs, on average, and include a small fraction of stars which
could not be identified as belonging to any sub-galactic system (neither
could their gaseous progenitors). We also checked for possible
contamination of the IHPs from very extended thick-disc populations, by
relaxing the condition for particles to be classified as part of the
disc components at $z=0$, assuming $J_{z}/J_{z, \rm max(E) } > 0.5$. No
significant differences were found. 

In the following sections we analyse the chemical and dynamical
properties of the IHPs and OHPs, as populated by contributions from the
three different sub-populations: debris stars, disc-heated stars, and
endo-debris stars. Table \ref{tab1} summarizes the percentage of stars
in each sub-populations and their median abundances. We emphasize that, at any given time (owing to their
extended orbits), stars that are members of the IHPs and OHPs can occupy
the same spatial regions (both the IHR and the OHR), with different
relative fractions at different distances from the centre.  We perform
an analysis and further discuss this important point, and its consequences,
in Tissera et al. (in preparation).

\section{The diffuse IHPs}

As discussed above, our simulated IHPs are complex structures, with
contributions from disc-heated stars, as well as from the debris and
endo-debris stellar sub-populations. The debris sub-population is the
primary contributor, with an average of $\sim 45 \%$ across the six
simulated haloes, while the endo-debris and disc-heated sub-populations
each contribute $\sim 25 \%$ to  $\sim 30 \%$ , on average.

\begin{figure*}
\hspace*{-0.2cm}\resizebox{5.5cm}{!}{\includegraphics{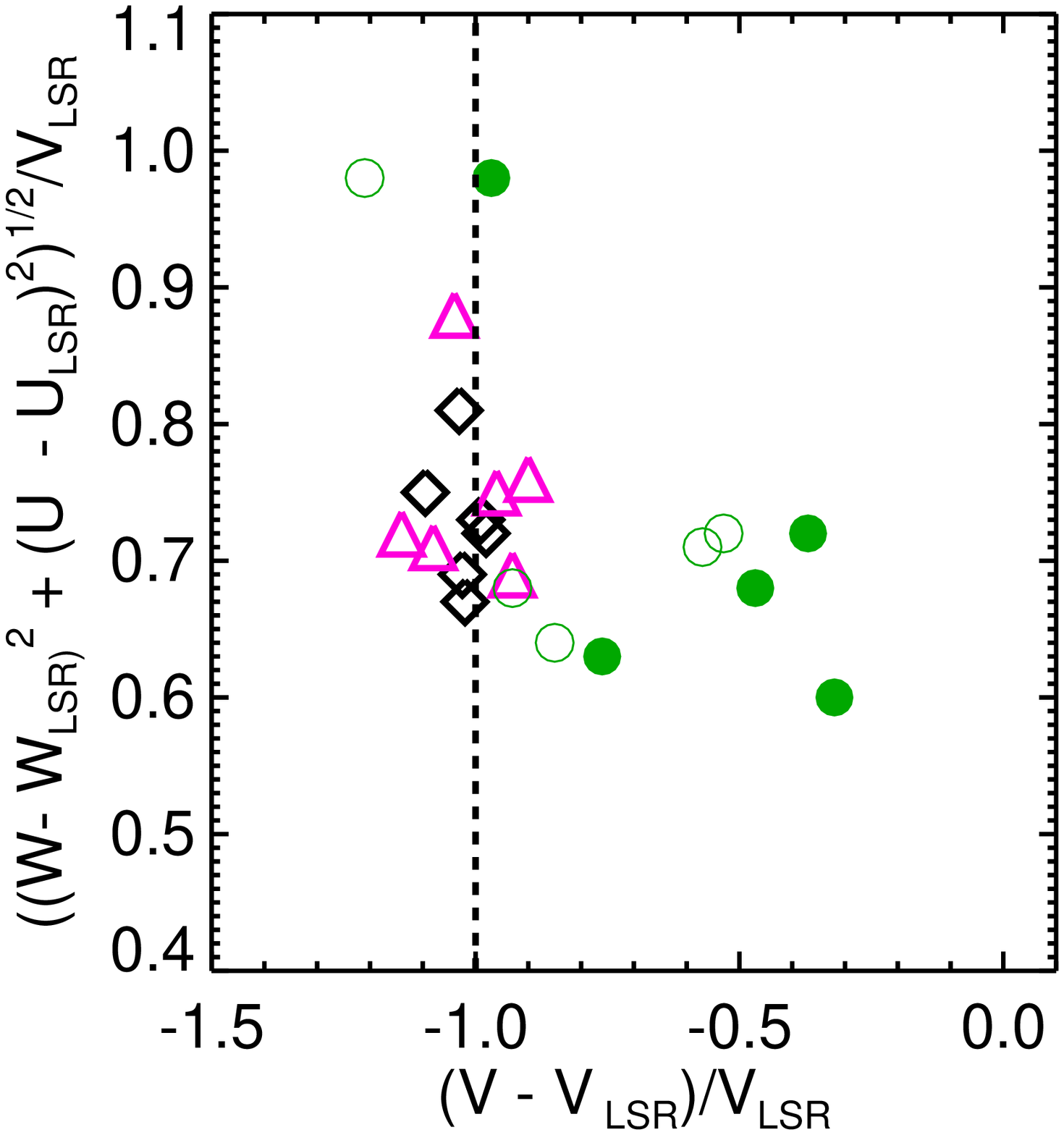}}
\hspace*{-0.2cm}\resizebox{5.5cm}{!}{\includegraphics{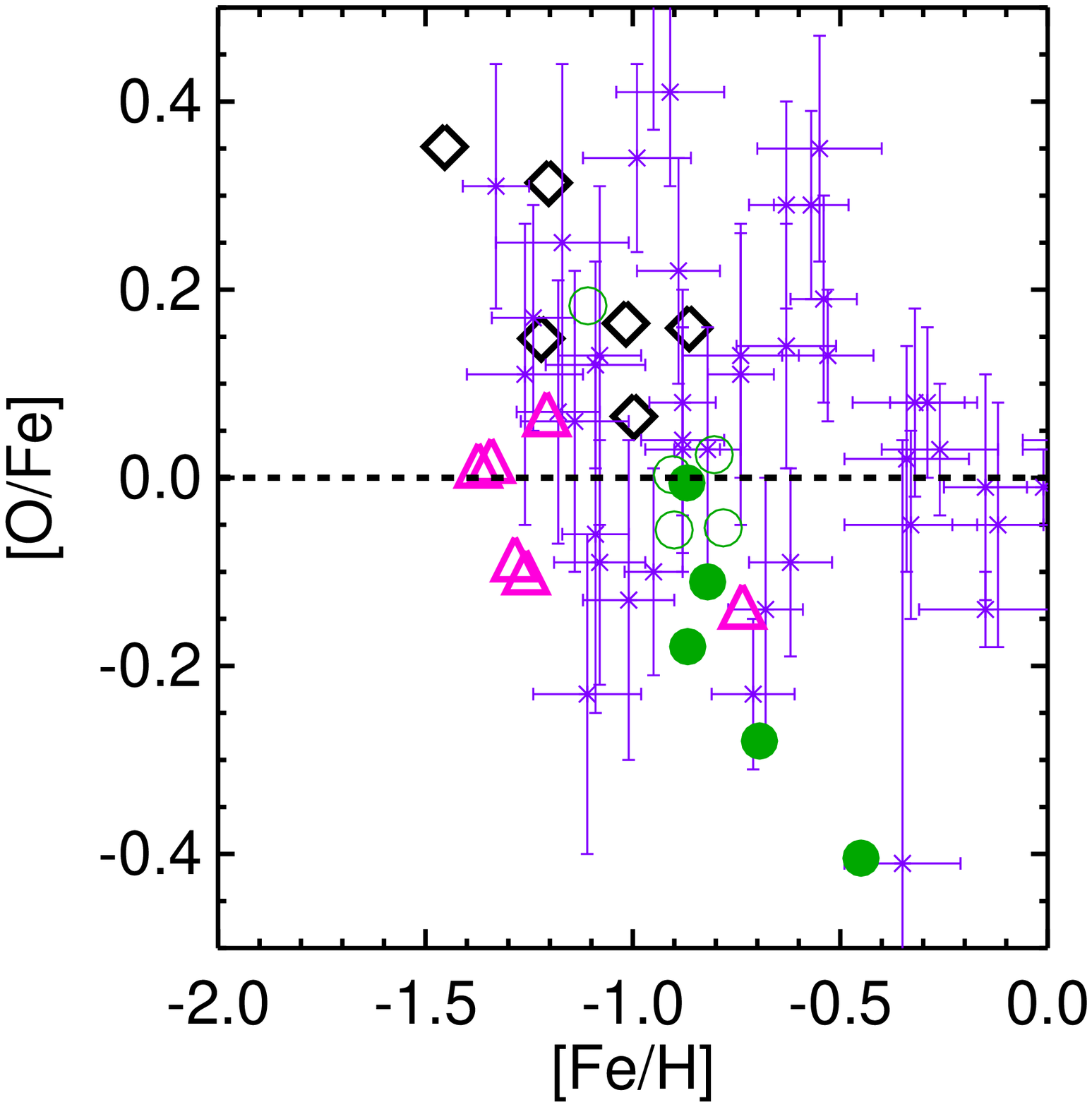}}
\hspace*{-0.2cm}\resizebox{5.5cm}{!}{\includegraphics{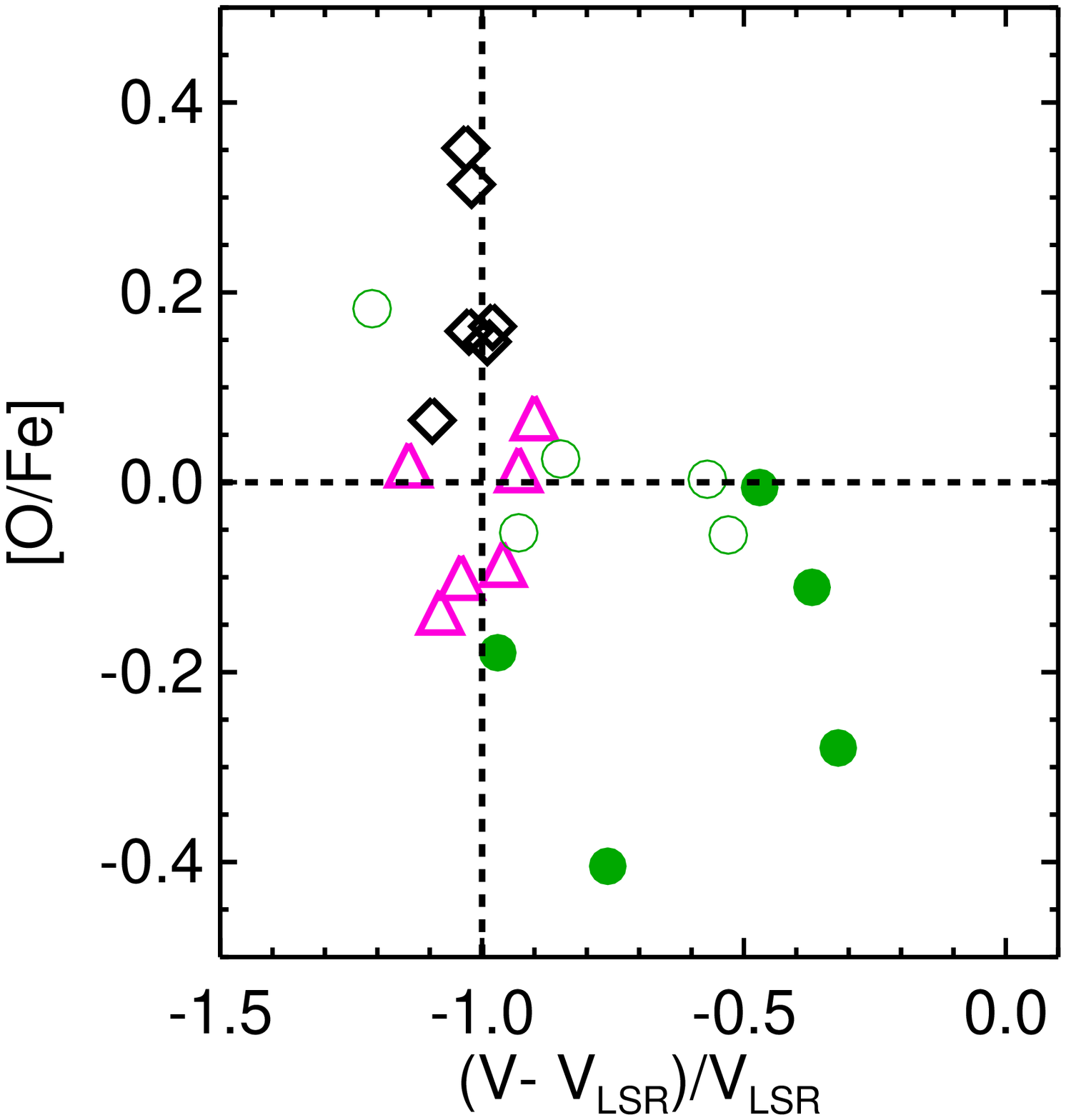}}
\caption{ Median rotation velocities for stars in the simulations assigned to the IHPs,
displayed as the Toomre diagram (left
panel), [O/Fe] versus [Fe/H] (middle panel), and [O/Fe] versus rotation
velocity for the averaged stellar sub-populations for each of the six
simulated haloes, encoded according to: debris stars (diamonds),
endo-debris stars (triangles), old disc-heated stars ($> 10$ Gyr; open
circles) and young disc-heated stars ($8-10$ Gyr; filled circles).
The vertical dashed lines in the left and right
panels denote the limit between prograde and retrograde rotation with
respect to the corresponding galactic disc in each simulation.  The
horizontal dashed lines in the middle and right panels indicate the
solar abundance ratio for [O/Fe]. Note that only five symbols are
included for the disc-heated stars since Aq-B-5 has a negligible
contribution of this sub-population. 
 In the middle panel, we show the
observations from \citet{sheffield2012} for a sample of individual
stars in the Solar Neigbourhood (violet points and errorbars). Note
that for the observed sample, each point represents the properties
of individual stars while
for the simulated data, median estimations over the sub-populations
are included. 
 }
\label{toomregeneral}
\end{figure*}

To better visualize the global differences between the IHPs in our
simulated galaxies, Fig. ~\ref{toomregeneral} (left panel) shows the
Toomre diagram defined by the median velocities, with each of the
contributing sub-populations marked.   Following \citet{mcc2012},
 the velocity coordinates $U, V$, and $W$ are estimated in the
  rest frame of fiducial Solar Neighbourhoods defined for  the stellar disc
  component of the simulated galaxies.
 The rest frames are defined by estimating
  the mean velocity components for stars in the stellar discs at
  $\approx 8$ kpc.  Velocities are normalized to the median rotation
  velocity of the disc, V$_{\rm LSR}$, in the fiducial Solar Neighbourhoods, 
in order to consider the
  differences in mass between the simulated galaxies. Hence, the limit
  between retrograde and prograde motions is set at $-1$.
 As can be seen from inspection of this figure, the
debris and endo-debris sub-populations exhibit, on average, little
evidence for strong rotation. Typically, disc-heated stars exhibit
different kinematics; most are more rotationally supported than the
stars in the other two sub-populations.  However, there exist an
important subset of disc-heated stars
which cannot be distinguished from debris or endo-debris stars. In fact,
there is a clear age dependence on the behaviour of the disc-heated
stars -- old disc-heated stars ($> 10$ Gyr; open circles) do not show
significant rotational velocities, while  younger stars ($< 10$ Gyr; filled circles)
exhibit clear signals of rotation. Note that, in these
simulations, young disc-heated stars are nevertheless older than $\sim
8$ Gyr \footnote{It is worth mentioning that the simulated discs at $z
  =0$ are
  dominated by young stars as shown by \citet{tissera2012}. However,
  discs contributed with heated stars to the IHPs  mainly
  at $z > 1$. For this reason, the disc-heated stars found
  in the IHPs are older than $\sim 8 $ Gyrs.}. In the case of Aq-A-5, a fraction of old disc-heated stars
actually exhibit retrograde rotation, illustrating the complexity and
diversity of halo assembly in a hierarchical clustering scenario.

Due to the time delay between the corresponding evolution of the
progenitor stars of  SNII and SNIa, the $\alpha$-element-to-iron ratios provide
information about the star-formation history of the systems. Fig.
~\ref{toomregeneral} (middle panel) shows the median [O/Fe] vs. [Fe/H]
for the IHPs, segregated into the different stellar sub-populations. On
average, debris stars are $\alpha$-enriched relative to iron.
However, we also find a number of stars with low [$\alpha$/Fe] ratios,
varying from $7\%$ to $35\%$, evidencing contributions from SNIa. The
fraction of such stars correlates with the mass of the accreted
sub-galactic systems, so that the larger fractions are found in  IHPs
that formed from more massive systems. The endo-debris and disc-heated
sub-populations have a larger fraction of low (or even sub-solar)
[$\alpha$/Fe] ratios. The stellar sub-populations also occupy different
locations in the  [O/Fe] vs [Fe/H] plane. Debris stars are metal poor, and have the
smallest contribution from low-[$\alpha$/Fe] stars. Endo-debris stars
have lower mean [$\alpha$/Fe] ratios. The disc-heated stars tend to
exhibit, on average, both higher metallicities and lower [$\alpha$/Fe]
ratios compared to the other sub-populations. As expected, this is
particularly true for the younger disc-heated stars.

We found that, on average, $55\%$ of the $\alpha$-enriched stars in IHPs
belong to the debris sub-components, while disc-heated stars and
endo-debris stars contribute, individually, on the order of $20-25\%$ of
the total mass fraction of $\alpha$-enriched stars, on average.
Observationally, there is evidence for the existence of halo stars with
low $\alpha$-enrichment \citep{nissen2010, sheffield2012}, although
their frequency is not yet clearly determined. The endo-debris
sub-populations in the IHPs could be associated with the low-$\alpha$ and
low [FeH] stars classified as the accreted stars by
\citet{sheffield2012}.  In the middle panel of
  Fig. ~\ref{toomregeneral}, we show the observations of  \citet{sheffield2012}
  superposed with the median values estimated for the three
  sub-populations in our  simulated IHPs\footnote{The differences in the estimators used for observations and simulations
 could explain in part the narrow range of metallicities covered by the
simulated data compared to the observations in this figure. The lack of very high metallicity stars in
the simulations could be also related to the action of very efficient mass-loaded simulated galactic winds 
 or to  a different evolutionary 
path for the MW compared to those of our simulated systems. It is beyond the scope of this paper to carry out a detailed comparison with the MW.}.
Observations of the [$\alpha$/Fe] abundance ratios for large, unbiased
samples of MW halo stars will provide very important constraints on
galaxy formation models. According to our present models, such
measurements would be a probe of the complex mixture of stellar
sub-populations in the inner haloes of large spiral galaxies such as the
MW and M31. Testable predictions of various formation and evolution
scenarios could then be made from which models can be revised and
improved. 

With respect to the mean rotational velocities, the $\alpha$-enriched
stars followed the same mean trends of their corresponding
sub-populations -- those belonging to the debris or endo-debris
sub-populations exhibit little or no net rotation with respect to the
galactic centre in the range $ [-15,15]$ km s$^{-1}$, while those
classified as disc-heated stars tend to exhibit clear prograde rotation
($\sim 50$ km s$^{-1}$), particularly the younger stars. The
$\alpha$-rich stars tend to be $1-2$ Gyr older than $\alpha$-poor stars.
A portion of the latter coming from the disc components can be up to 1
Gyr younger.

\begin{figure*}
\hspace*{-0.2cm}\resizebox{4.5cm}{!}{\includegraphics{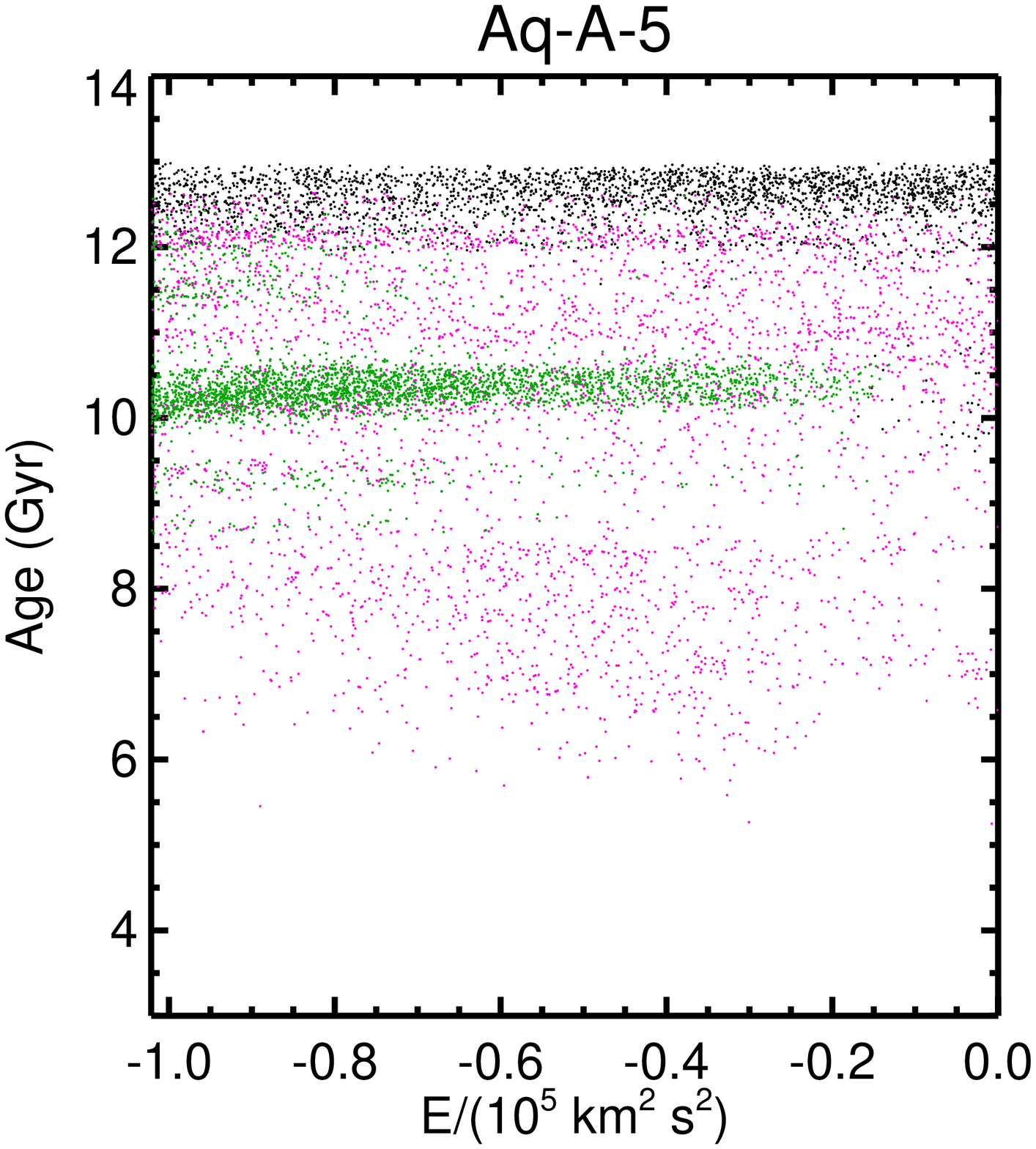}}
\hspace*{-0.2cm}\resizebox{4.5cm}{!}{\includegraphics{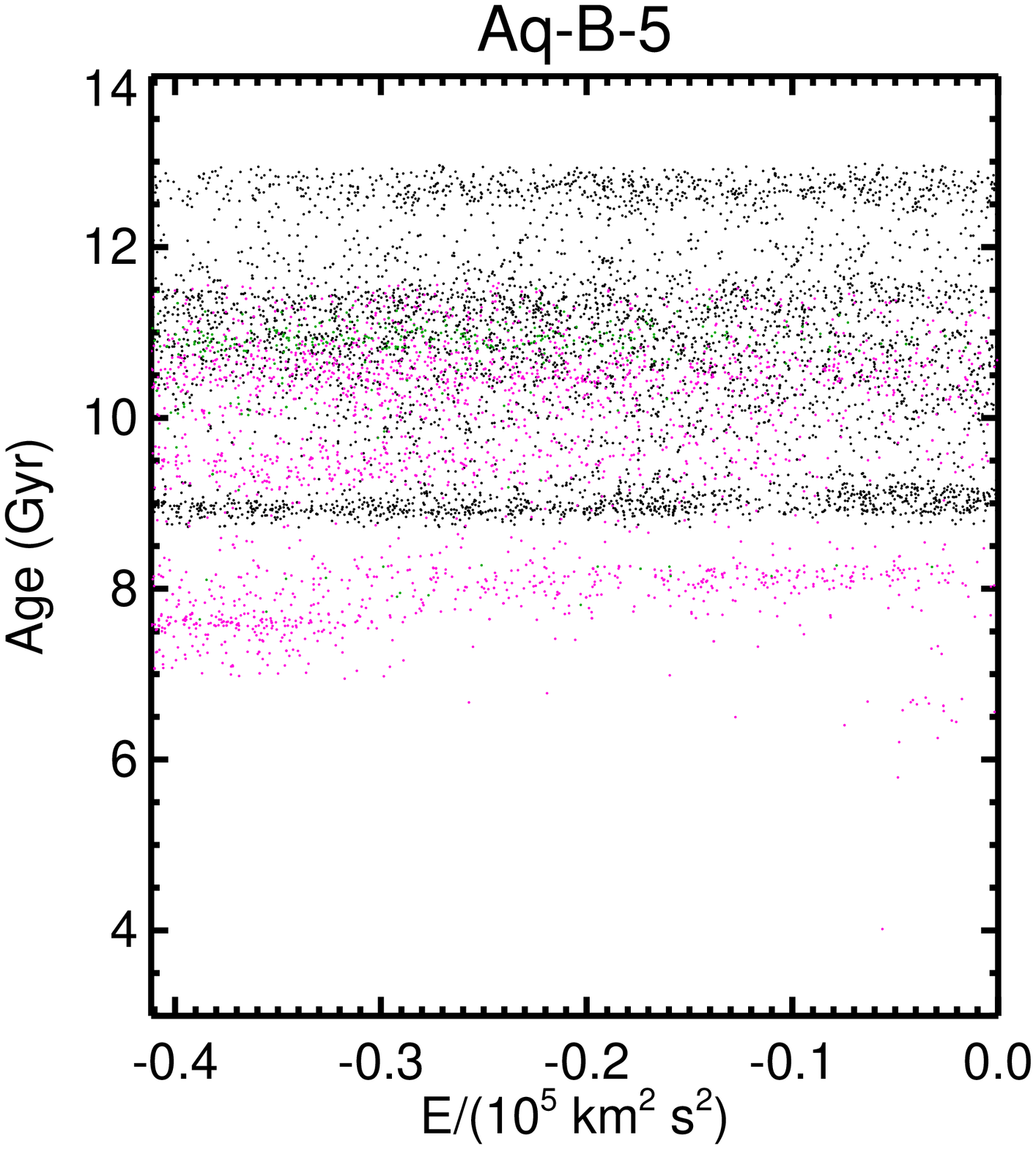}}
\hspace*{-0.2cm}\resizebox{4.5cm}{!}{\includegraphics{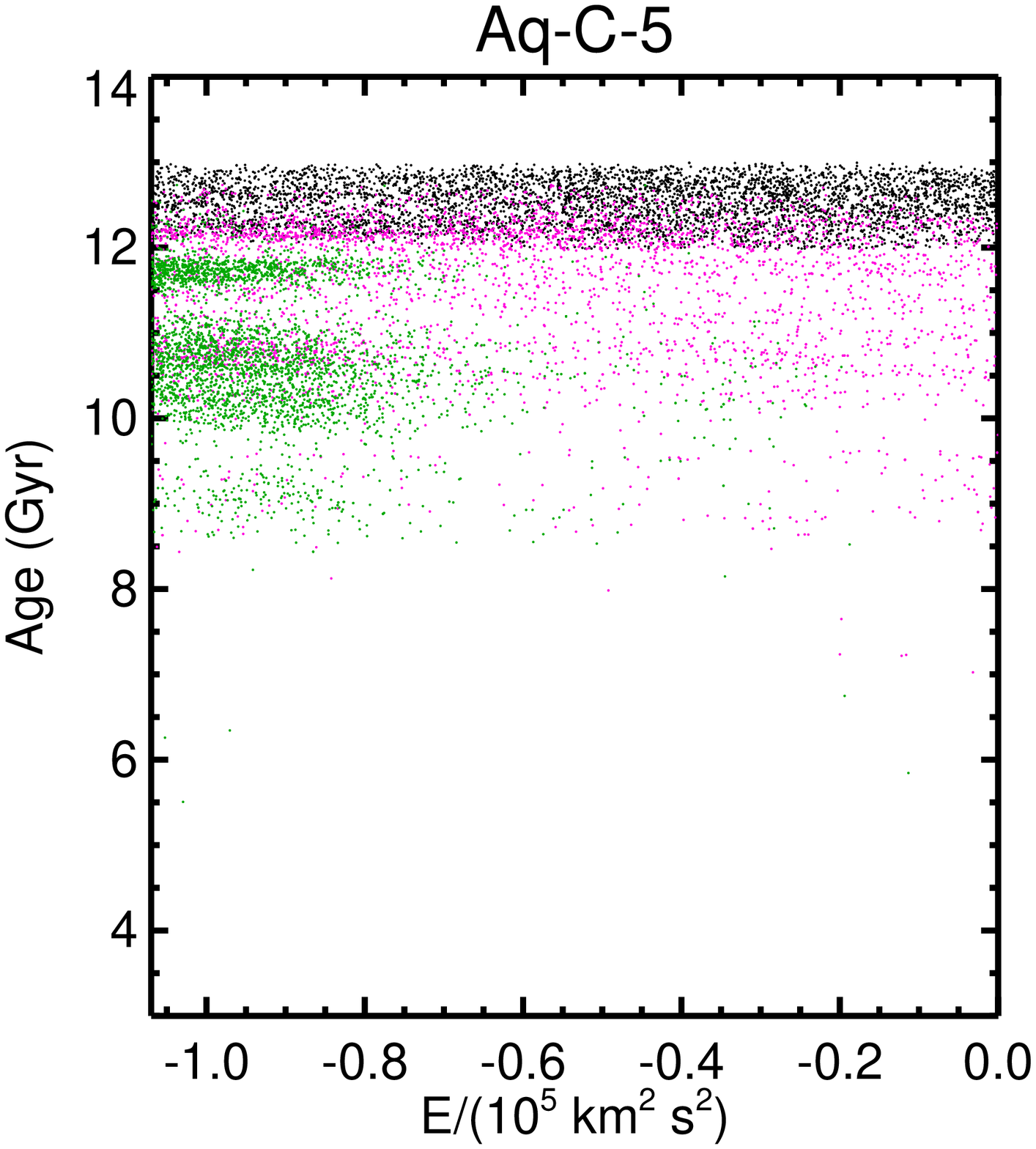}}\\
\hspace*{-0.2cm}\resizebox{4.5cm}{!}{\includegraphics{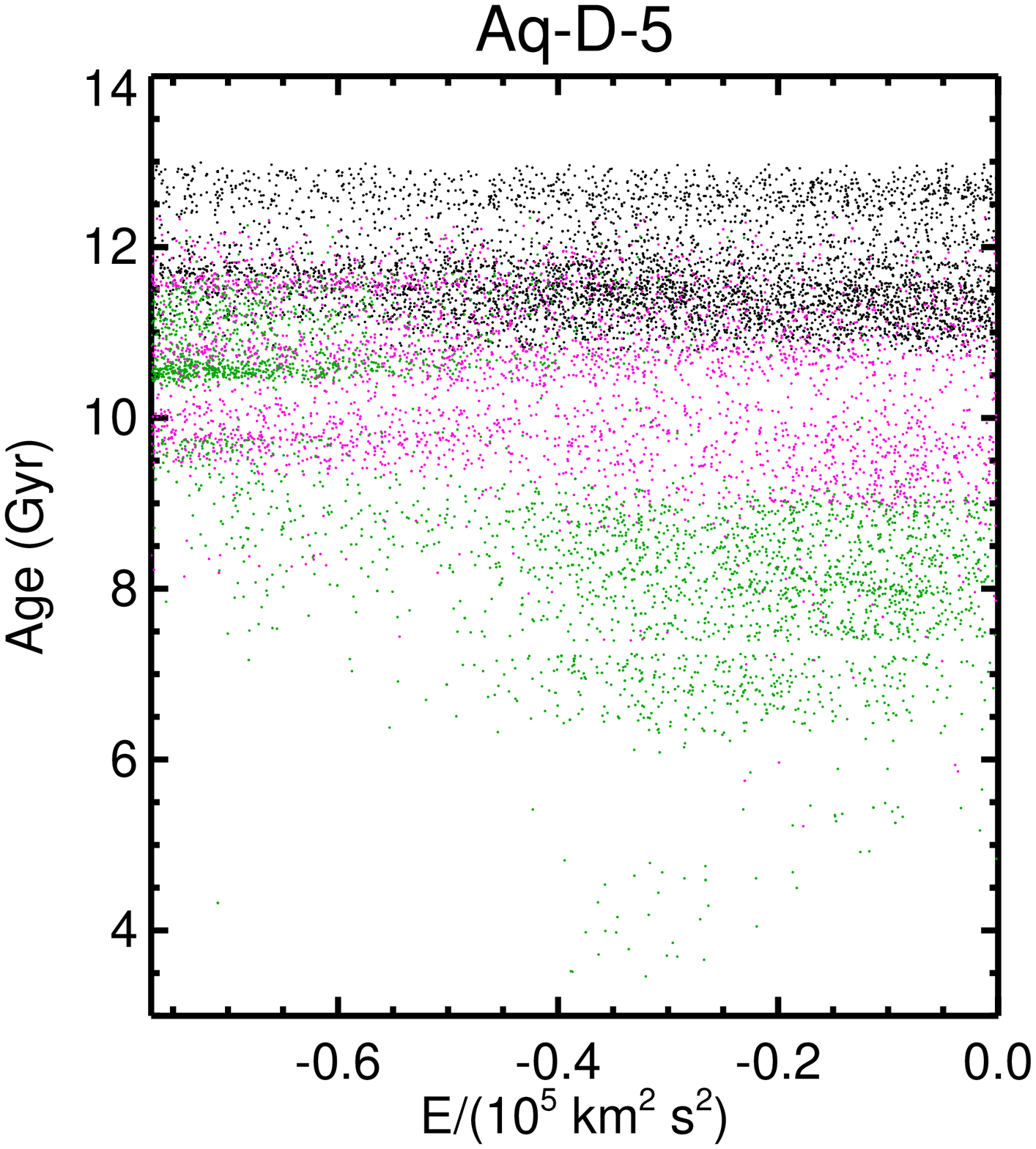}}
\hspace*{-0.2cm}\resizebox{4.5cm}{!}{\includegraphics{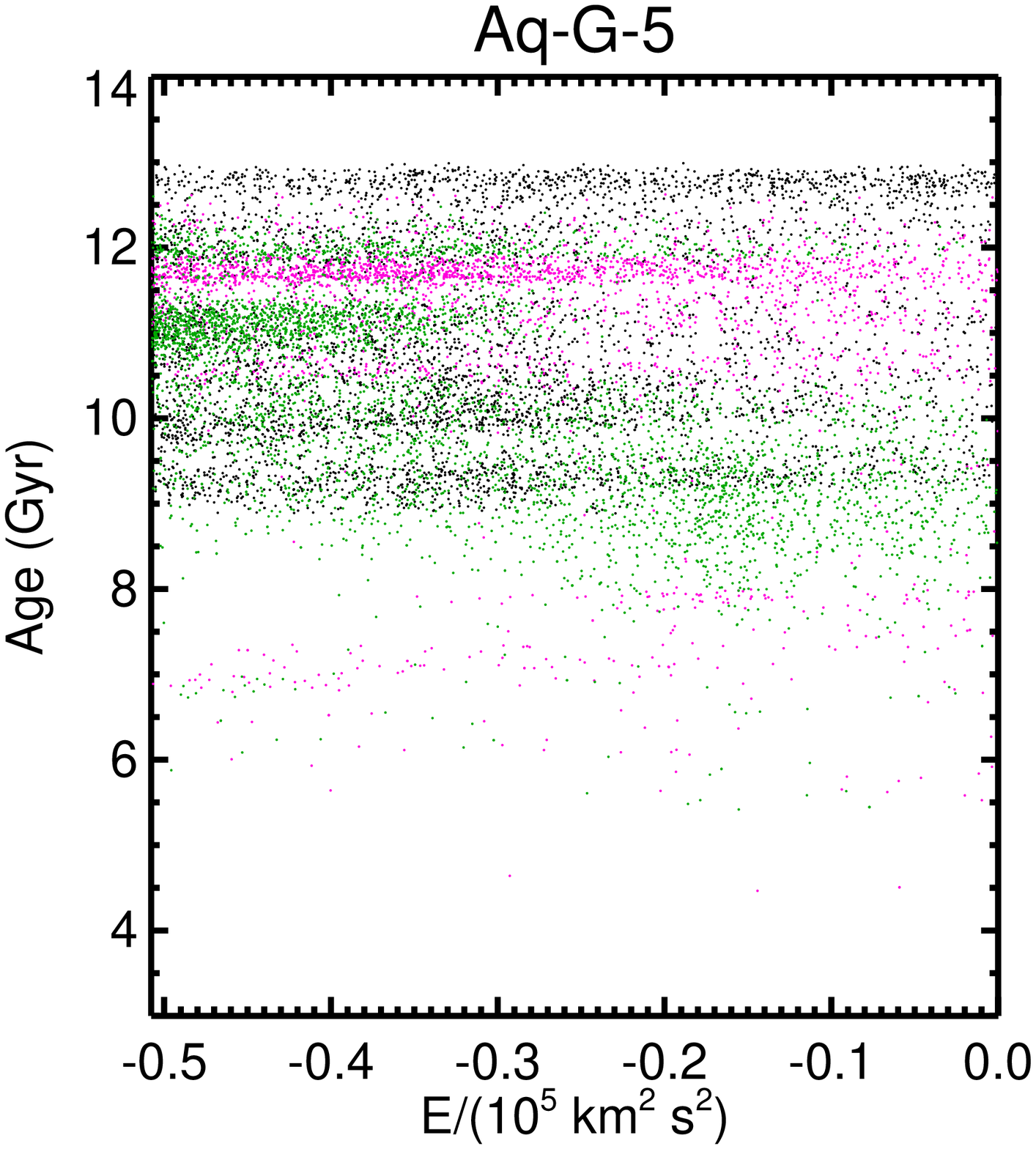}}
\hspace*{-0.2cm}\resizebox{4.5cm}{!}{\includegraphics{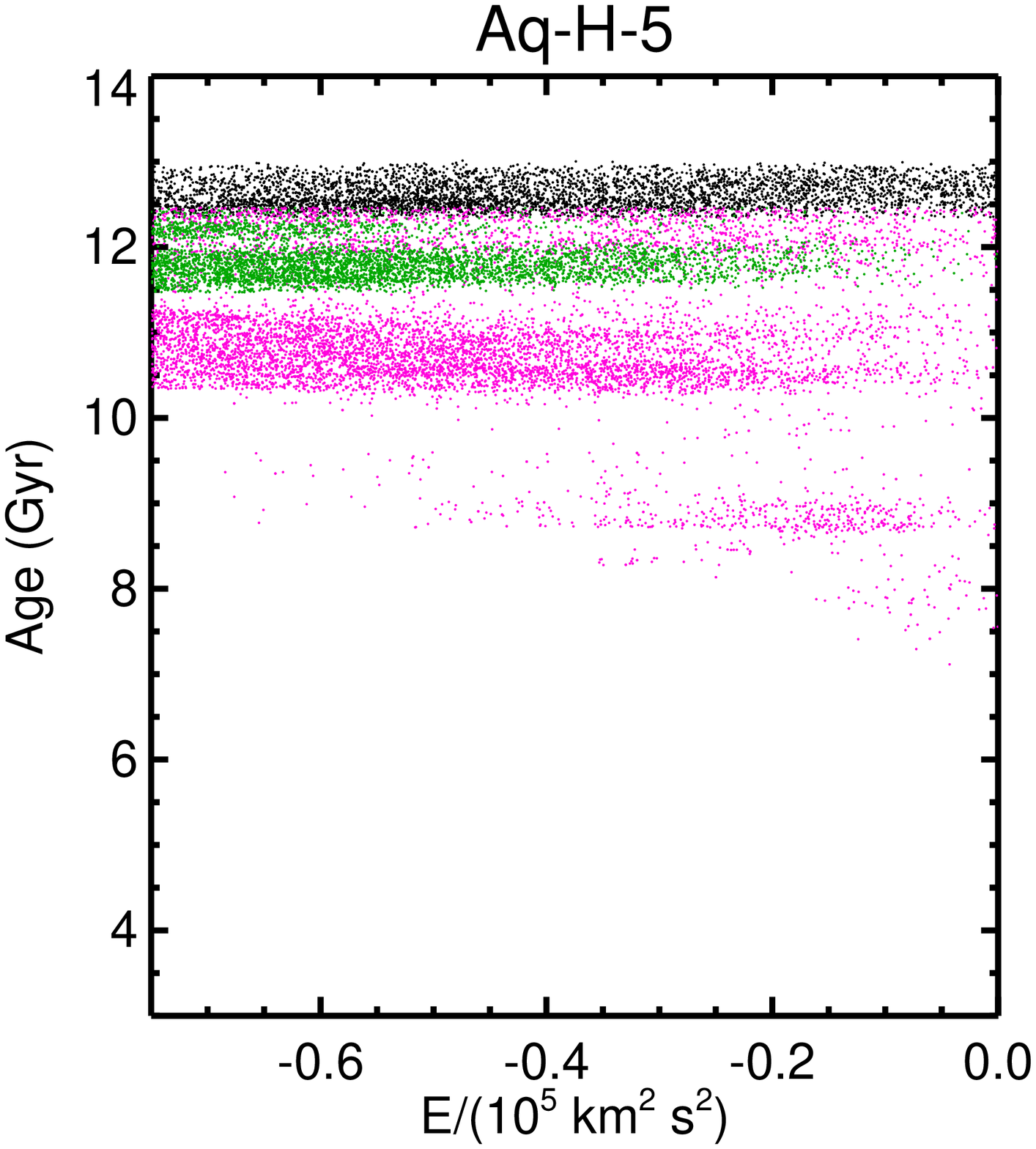}}
\caption{Age versus binding energy for the IHPs associated with each of
our simulated haloes. Stars
are coloured according to their assigned sub-population:
debris stars (black), endo-debris stars (magenta) and disc-heated stars (green).
For the sake of clarity, we only show $\sim 30\%$ of the total sample. }
\label{ages}
\end{figure*}

Fig. ~\ref{ages} shows the ages of the IHPs, as a function of
binding energy, for the three different stellar sub-components. The
debris stars are the oldest, and they span the entire range of total
energies. These stars are among the first ones to be formed; as a
consequence, they are primarily low metallicity and $\alpha$-enriched
relative to iron (as can be appreciated from inspection of
Fig.~\ref{toomregeneral}). The endo-debris stars tend to be slightly
younger, and relatively more gravitationally bound. They formed in
gas-rich satellites accreted at $z > 3$; $50\%$ of them were formed
in sub-galactic systems with dynamical masses smaller than $10^{8.5} $
M$_\odot$.  On average, these stars have a larger fraction of low
$\alpha$-enrichment when compared to the debris sub-populations. As
mentioned before, there are two distinctive regions occupied by
disc-heated stars -- the most tightly bound stars tend to be older and
$\alpha$-enriched, while the younger disc-heated stars extend to higher
energies and are more $\alpha$-poor. 

Our IHPs comprise primarily very old stars, with more than $50\%$ older
than 10 Gyrs (a lookback time comparable to redshift $z \approx 2$),
while, on average, only $\sim 5\%$ of them are younger than $\sim 7.5$
Gyrs ($z \approx 1$). Most of the latter stars are disc heated. As
mentioned above, the oldest stars tend to dominate at higher energies,
thus the proportion of old and young stars might be expected to vary
with galactocentric distance. Observations of stellar populations in
galactic haloes at higher redshifts should help to constrain their
assembly \citep{trujillo2012}.

\subsection{[Fe/H] profiles of the diffuse IHPs}

The different chemical abundances and binding energy distributions of
the stellar sub-populations can imprint differences onto their
metallicity distribution profiles, which, in turn,  provides information
on the relative contributions of each of the sub-populations.
Previous works have examined the metallicity profiles of their simulated
stellar spheroids as a whole (i.e., central spheroids, inner and outer
stellar haloes; Font et al. 2011), or looked at the difference between
the inner and the outer stellar haloes \citep{tissera2012}, and reported
that these simulations, within the concordance $\Lambda$CDM scenario,
are able to reproduce metallicity trends similar to those reported by
observers for the MW and M31 (see Introduction for a discussion on this
point).

\begin{figure*}
\hspace*{-0.2cm}\resizebox{4.5cm}{!}{\includegraphics{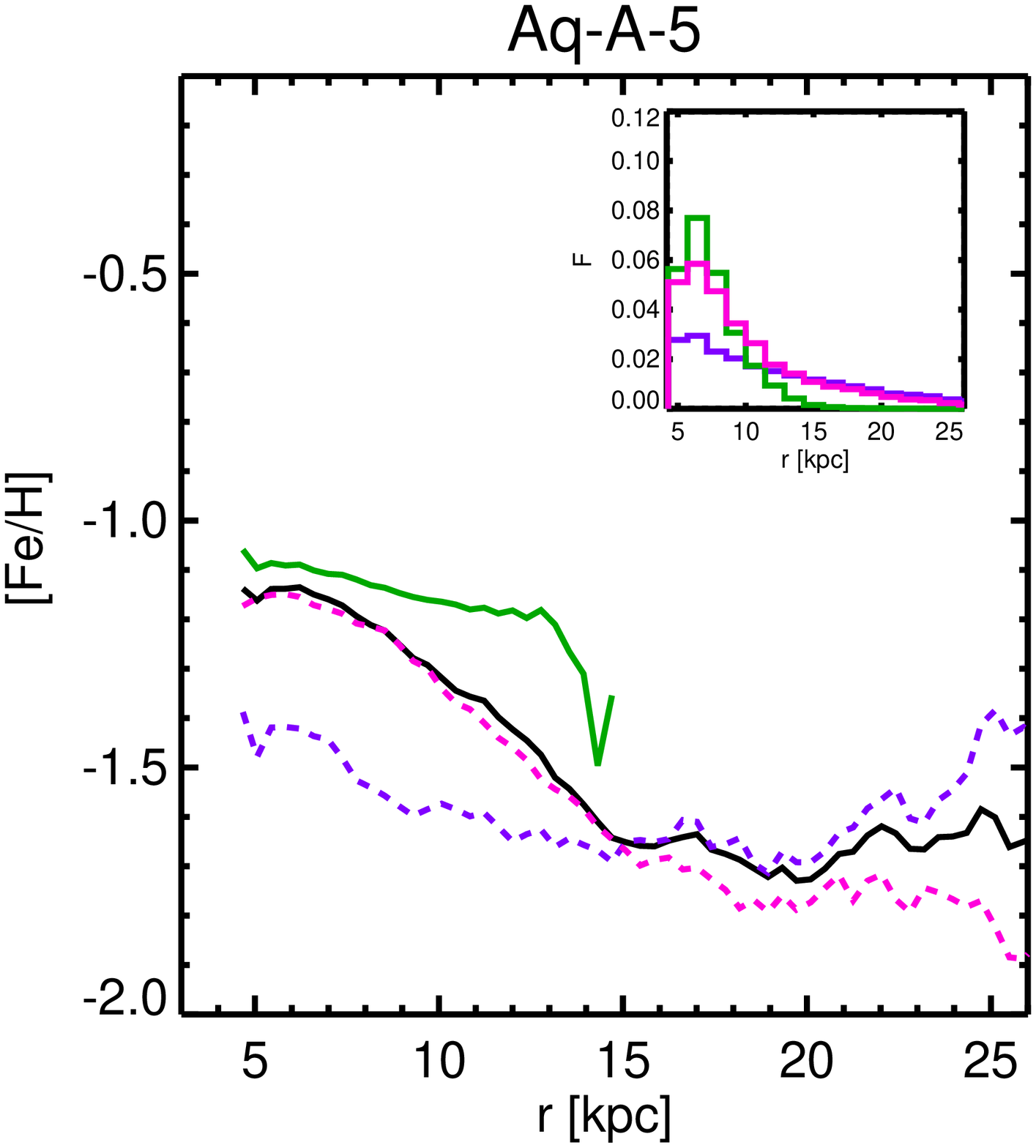}}
\hspace*{-0.2cm}\resizebox{4.5cm}{!}{\includegraphics{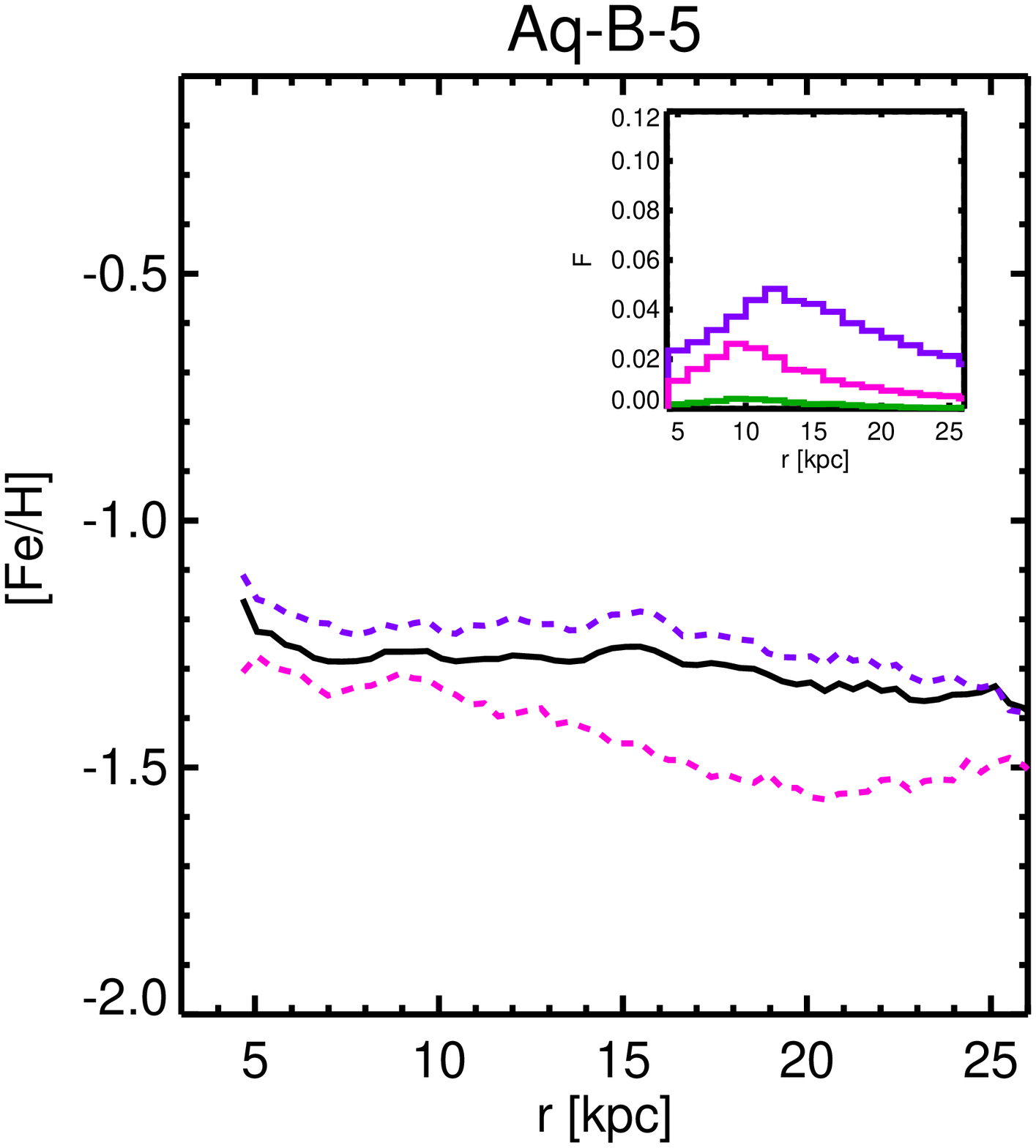}}
\hspace*{-0.2cm}\resizebox{4.5cm}{!}{\includegraphics{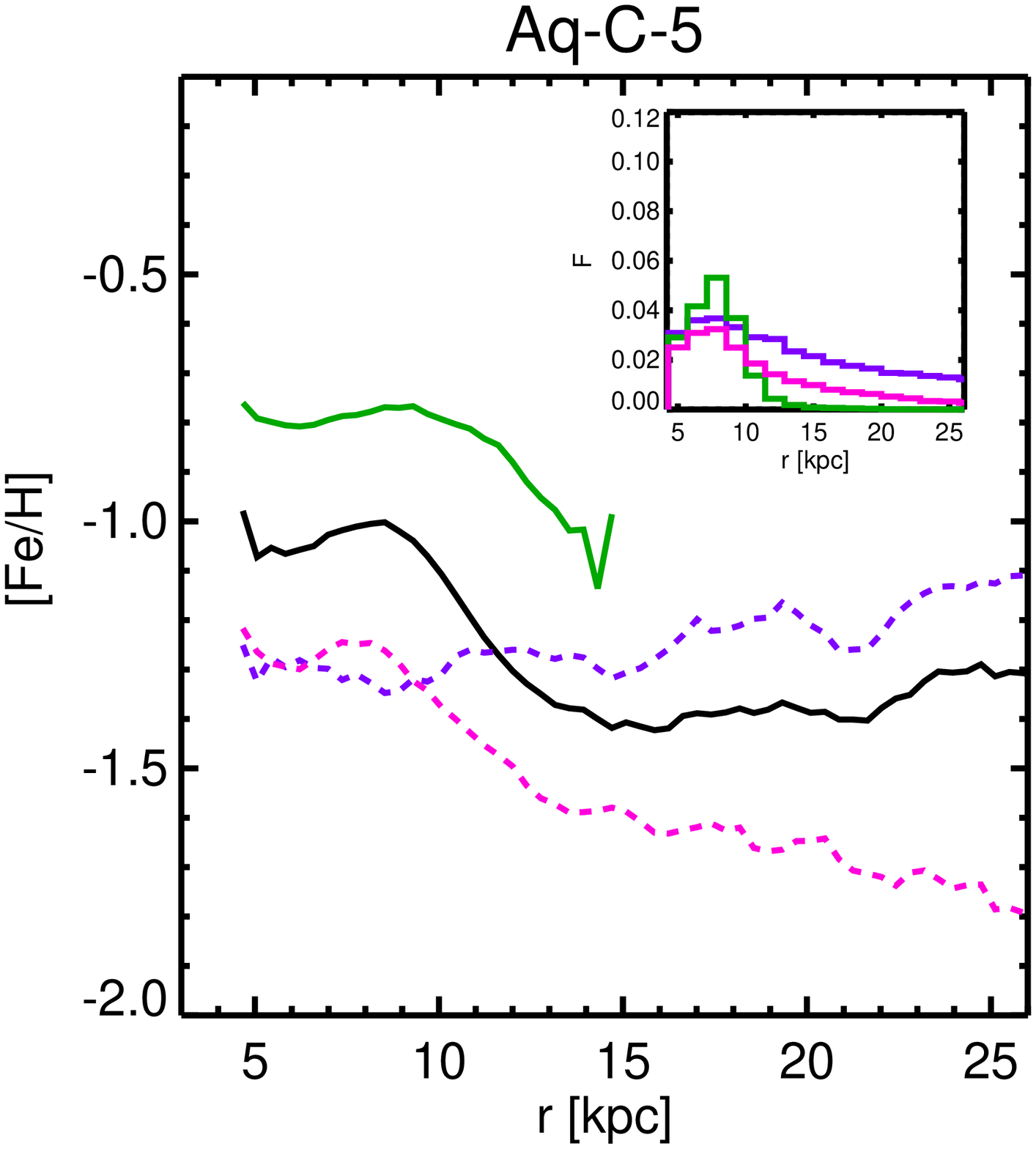}}\\
\hspace*{-0.2cm}\resizebox{4.5cm}{!}{\includegraphics{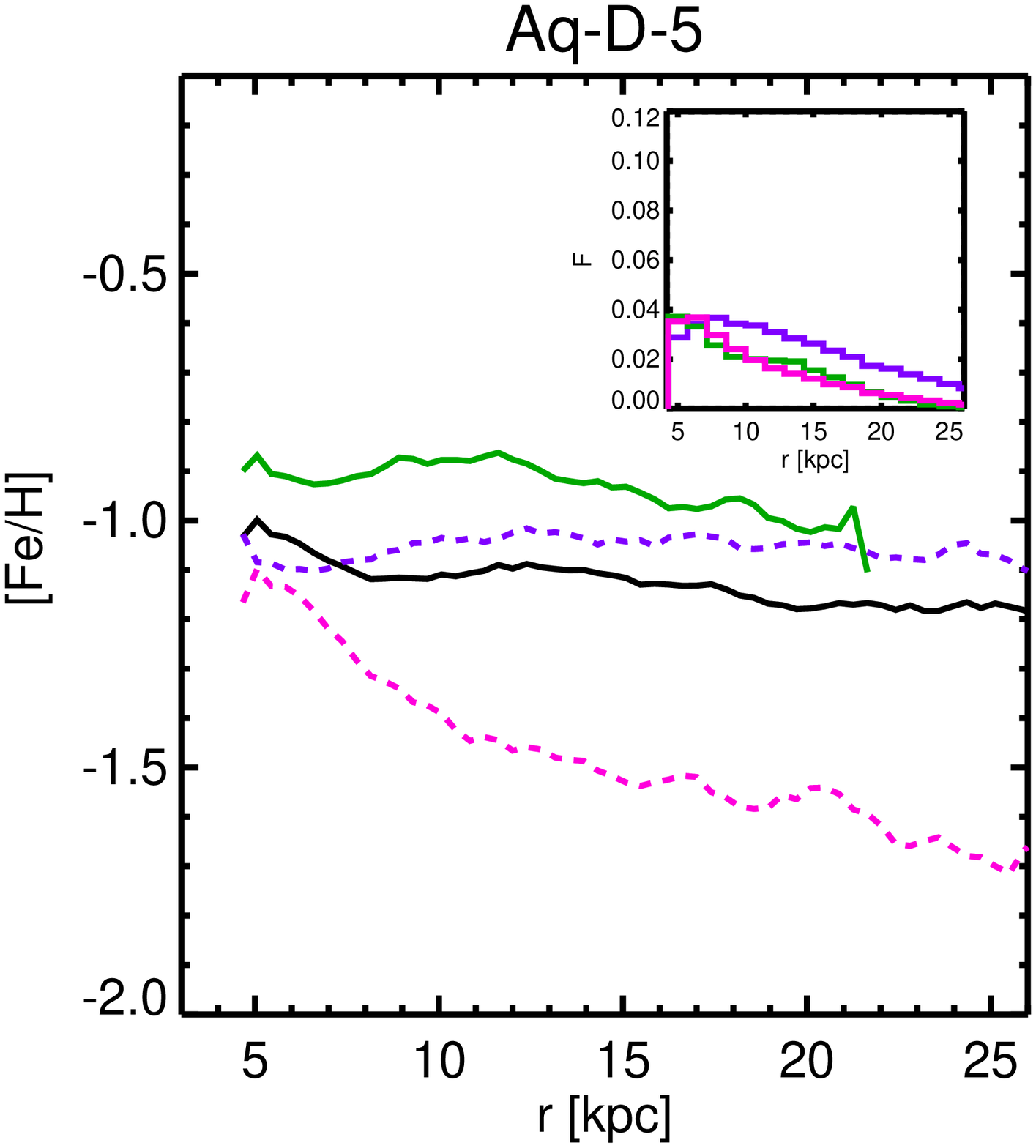}}
\hspace*{-0.2cm}\resizebox{4.5cm}{!}{\includegraphics{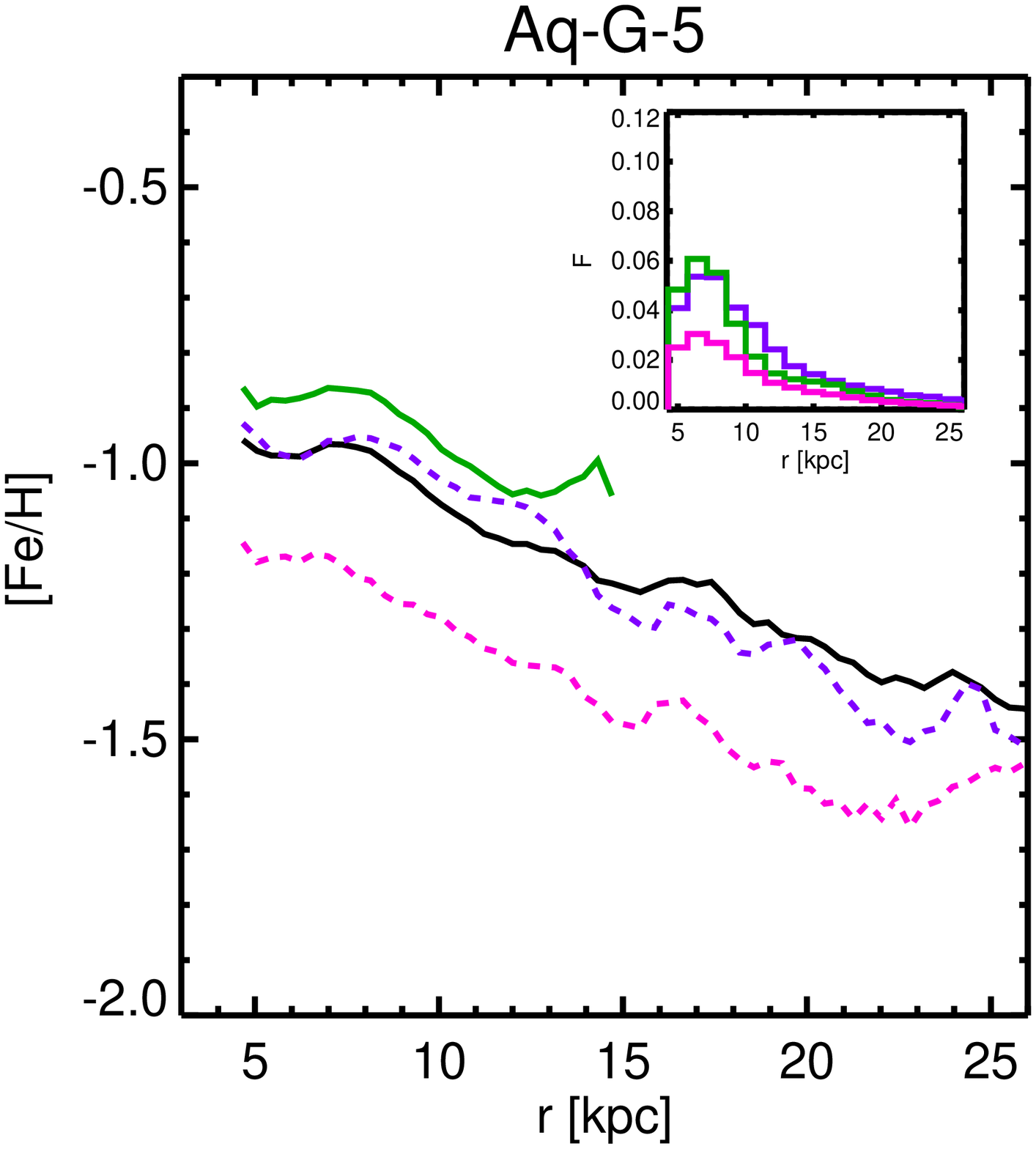}}
\hspace*{-0.2cm}\resizebox{4.5cm}{!}{\includegraphics{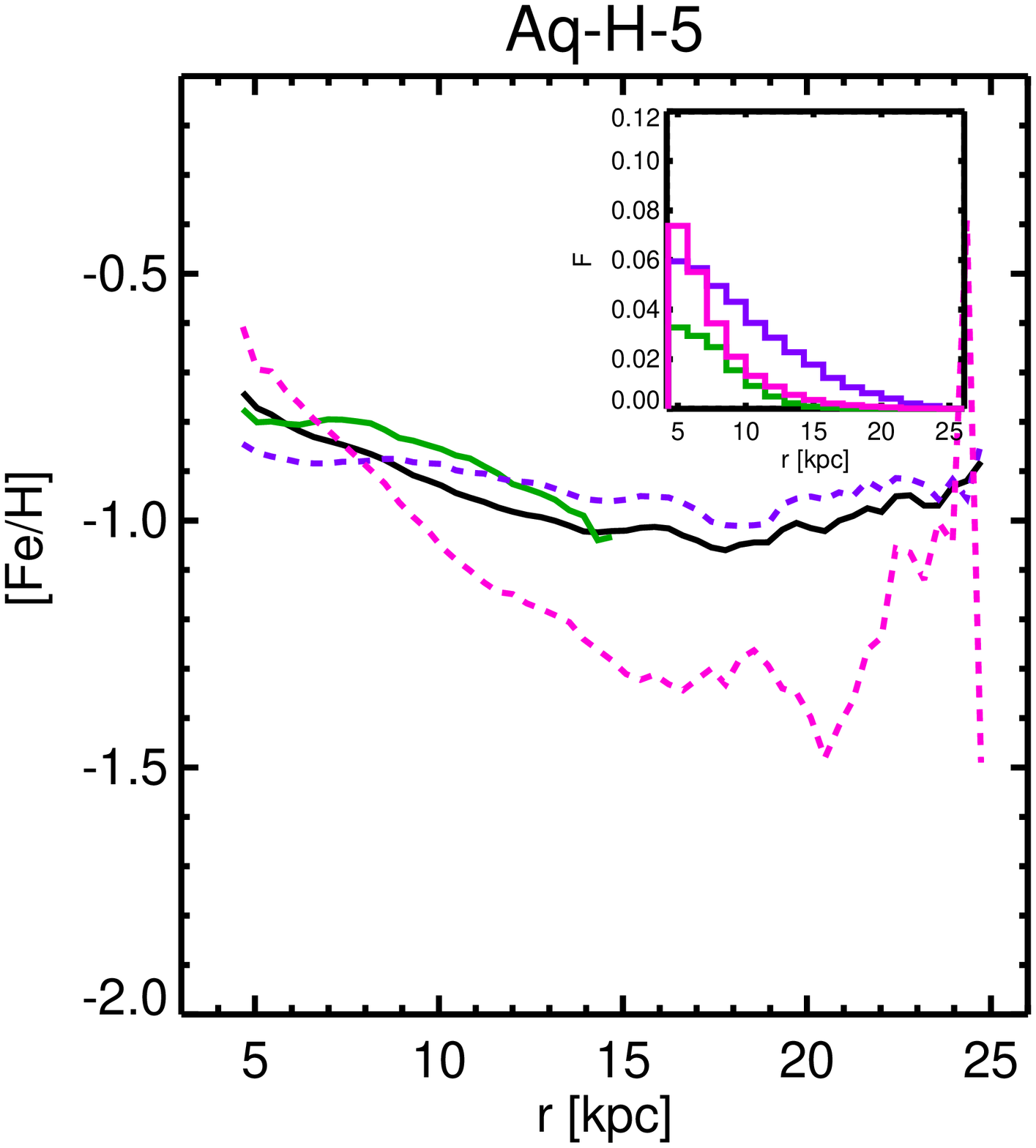}}
\caption{Abundance profiles for stellar
sub-populations assigned to the IHPs of our six simulated haloes
(black lines). The so-called \insitu component, made up of disc-heated stars  (green lines)
and endo-debris stars (magenta lines),
tends to be located in the central regions and to be dominated by low
binding energy particles. The debris stars (violet  dashed lines)  are more uniformly distributed with
distance, and as a consequence, dominate at larger radii. The total
abundance profiles are shown in black lines. The inset shows the
spatial distribution of the three sub-populations. 
 }
\label{gradhaloes}
\end{figure*}

Fig. ~\ref{gradhaloes} shows the [Fe/H] profiles of the disc-heated,
endo-debris, and debris stars, as well as the total
profiles. The abundance profiles represent the median [Fe/H] estimated  in equal-size radial
intervals (displayed medians have been estimated with at
least 100 stellar particles). The total metallicity
profile is clearly a combination of the contributions from each of the
sub-populations, the relative importance of which varies from
galaxy-to-galaxy. The resulting metallicity profiles depend on both the
level of enrichment of the different stellar sub-populations, and their
final spatial density distributions. The latter has been included in the
corresponding insets of Fig. ~\ref{gradhaloes}. Hence, although each of
these sub-populations may exhibit weak gradients individually, when they
are combined, the outcome could be a steeper profile. In fact, in some
cases, there exists a sharp transition between the region dominated by
the \insitu sub-populations (i.e., disc-heated plus endo-debris stars)
and that dominated by the debris sub-population. In such cases, fitting a
simple linear relation might provide spurious results. Consequently, we
prefer to understand the characteristics of the metallicity profiles by
analysing the spatial and binding energy distribution of each stellar
sub-population. This allows us to assess the role played by each
sub-population in shaping the final abundance profiles of the IHPs. The
eventual aim is to make use of such profiles (or quantities derived from
them) as tools for describing the assembly histories of individual
galaxies. 

From inspection of Fig.~\ref{gradhaloes}, there are three IHRs that
exhibit steeper variations of [Fe/H] with distance than the rest. In
these cases, the \insitu sub-components (disc-heated stars and/or
endo-debris stars) are more centrally concentrated with respect to the
debris sub-populations. In the cases with flat [Fe/H] profiles, the
three stellar sub-populations are better mixed. This can be seen from
the inset of Fig. ~\ref{gradhaloes}, where we display the spatial
distributions of the three different stellar sub-populations in the six
analysed haloes. It is clear that Aq-A-5, Aq-C-5, and Aq-G-5 have their
\insitu stars more concentrated toward lower energies (see Fig.
~\ref{ages}) and the central regions with respect to the debris stars.
Since the disc-heated stars and, in some cases, the endo-debris stars,
have higher metallicities in the central regions than debris stars, the
abundance profiles are steeper. It can be also seen that the debris
stars dominate at larger distances.

According to our simulations, a steep [Fe/H] profile is the result of
poorly mixed sub-populations, with significant contributions from either
disc-heated stars, endo-debris stars, or both, in the central regions.

\section{The diffuse OHPs}

The OHPs in our simulated halo systems comprise stars older than $\sim
11$ Gyr, which are very metal poor and $\alpha$-enriched relative to
iron. They formed from the accretion of satellite systems, so that most
of the stars belong to the debris sub-population. There is a small
fraction of stars that originated from an endo-debris sub-population
(which were in part classified as \insitu in Paper I; see see Section
2.2).  

\begin{figure*}
\hspace*{-0.2cm}\resizebox{5.5cm}{!}{\includegraphics{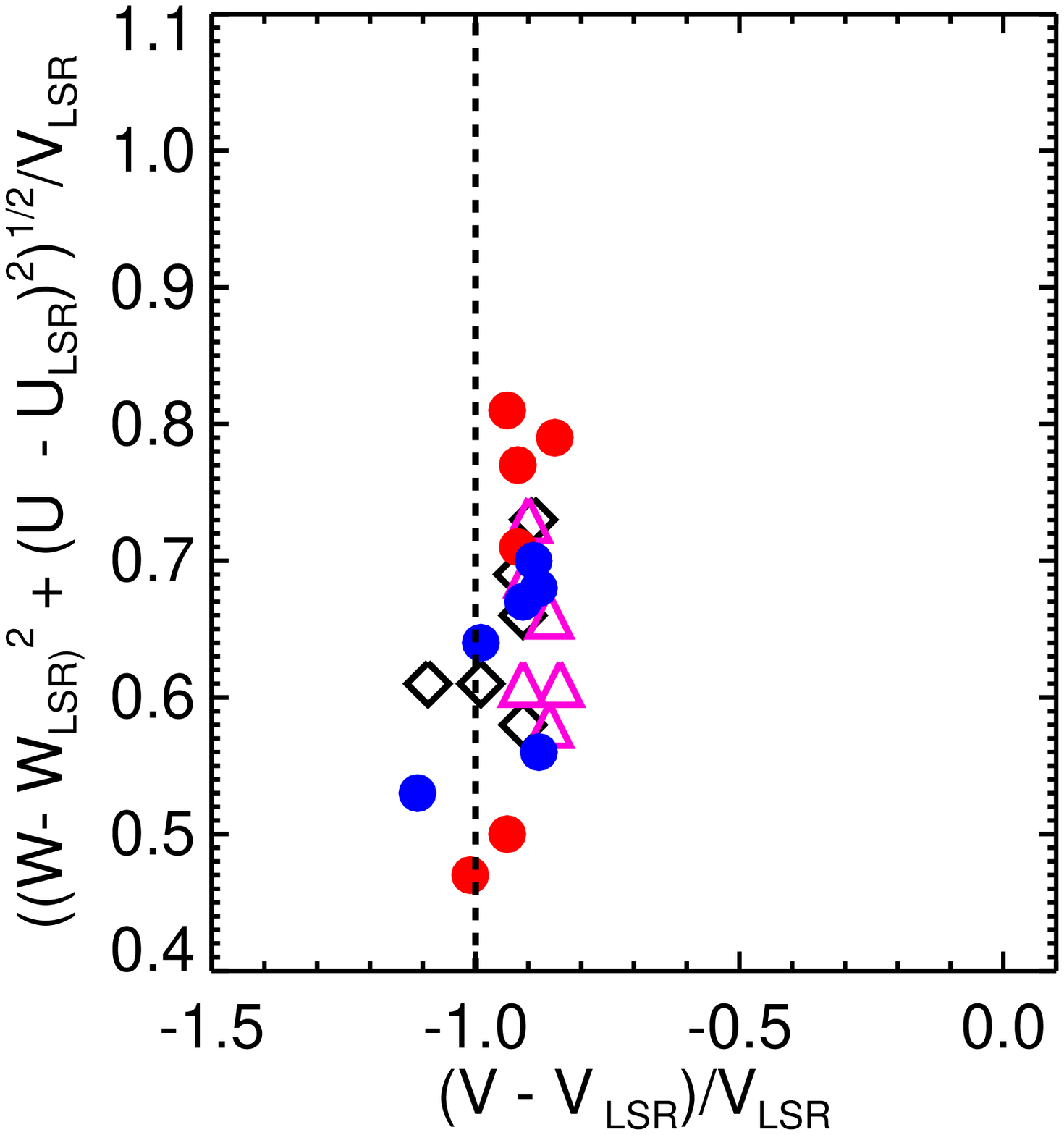}}
\hspace*{-0.2cm}\resizebox{5.5cm}{!}{\includegraphics{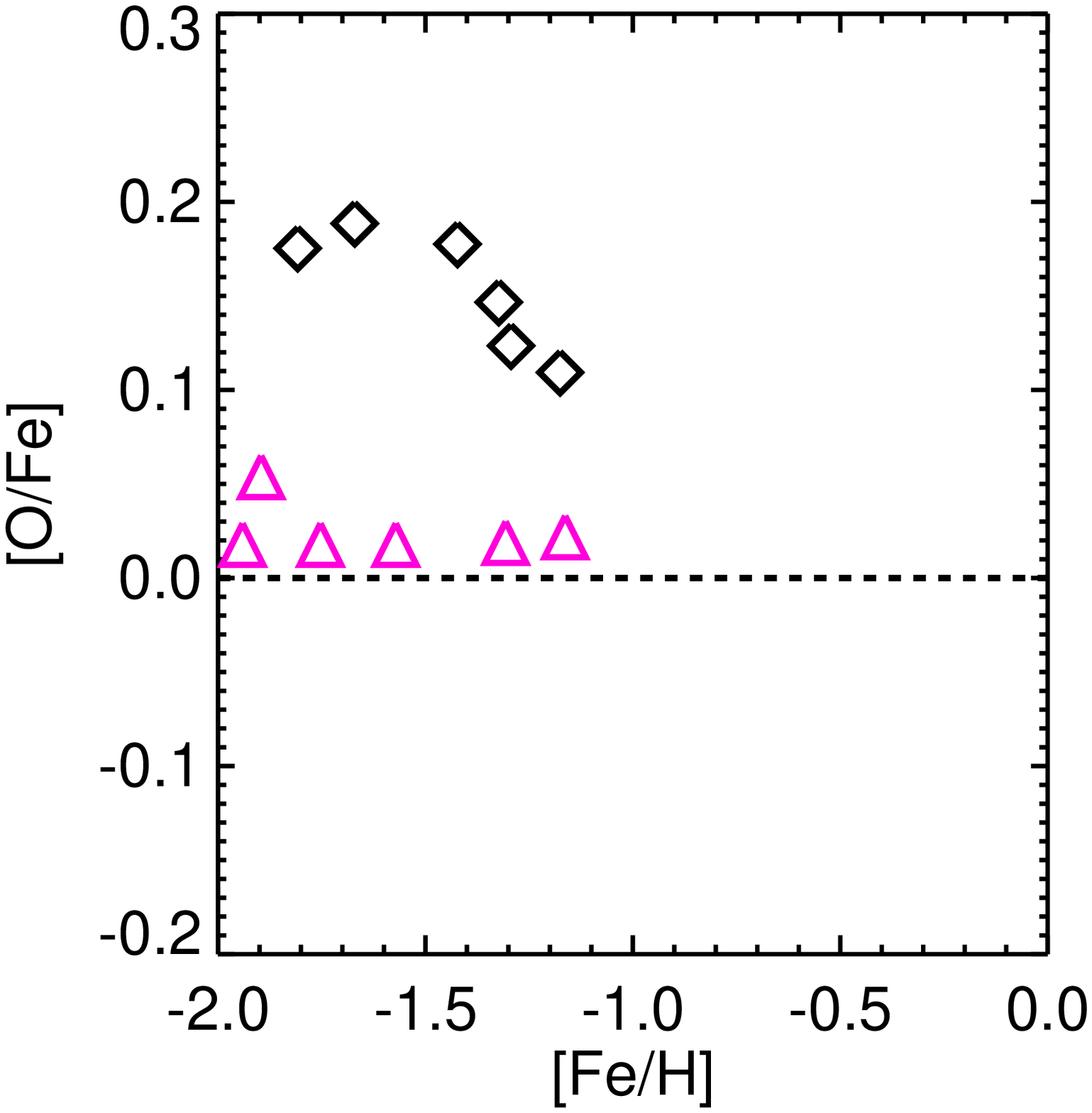}}
\hspace*{-0.2cm}\resizebox{5.5cm}{!}{\includegraphics{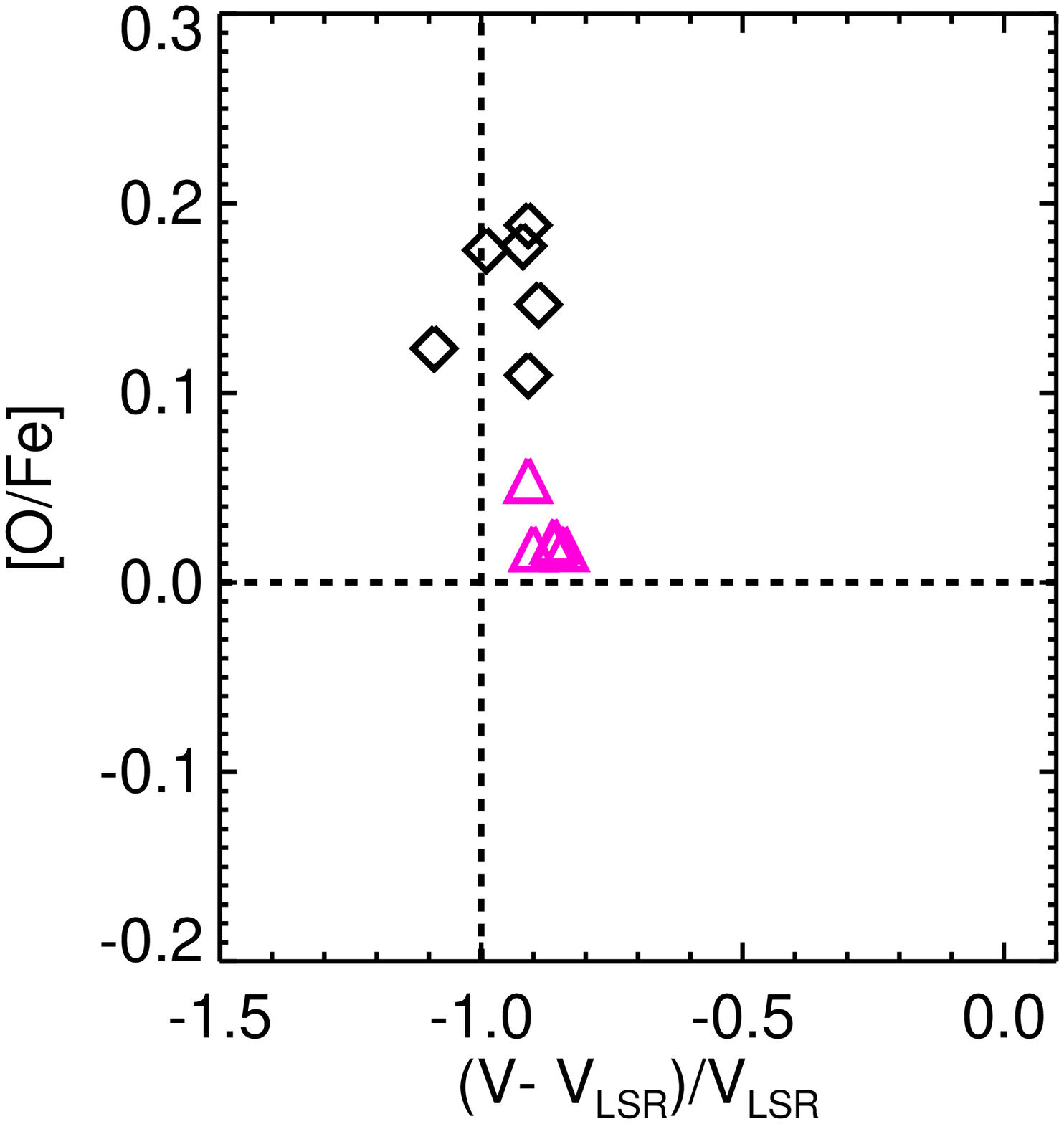}}
\caption{Median rotation velocities for stars in the simulations assigned to the OHPs,
displayed as the Toomre diagram (left
panel), [O/Fe] versus [Fe/H] (middle panel), and [O/Fe] versus rotation
velocity for the averaged stellar sub-populations for each of the six
simulated haloes, encoded according to: debris stars (diamonds) and
endo-debris stars (triangles). 
We also included the estimates for the high [$\alpha$/Fe]
stars (red circles) and low [$\alpha$/Fe] stars (magenta circles).
The
vertical dashed lines in the left and right panels denote the limit
between prograde and retrograde rotation with respect to the
corresponding galactic disc in each simulation. The horizontal dashed
lines in the middle and right panels indicate the solar abundance ratio
for [O/Fe].}
\label{toomrehaloex}
\end{figure*}

Regarding the kinematics, we find that most of the OHPs 
exhibit low net global rotation with respect to their corresponding galactic
centres, with mean values in the range $<V_\phi > \sim 20-25 $ km
s$^{-1}$. One of the six OHPs (Aq-G-5) has a mean $<V_\phi > \sim 7 $ km
s$^{-1}$, and another (Aq-C-5) exhibits a retrograde net rotation
$<V_\phi > \sim -33 $ km s$^{-1}$. At least in our simulations, the
stars are very well-mixed dynamically, so no significant differences in
the Toomre diagram could be found between the debris stars and the
endo-debris stars, or between the low- and high-[$\alpha$/Fe]
sub-populations (left panel of Fig. ~\ref{toomrehaloex}).

The middle panel of Fig. ~\ref{toomrehaloex} shows the [O/Fe] versus
[Fe/H] plane for the debris and endo-debris sub-populations. It is very
clear that the endo-debris sub-populations exhibit a larger contribution
from low-[$\alpha$/Fe] stars, on average. However, there is no trend
with rotational velocity, as can be seen from the right panel of Fig.
~\ref{toomrehaloex}. The existence (or absence) of such stars might be
tested with observations; their frequency could provide strong
constraints on galaxy-formation models and on how the star-formation
activity proceeded in small systems at high redshift. Note that
most of the stars in the OHPs are located outside the Solar Neighbourhood; less
than $\approx 10\%$ can be found within this region (Tissera et al. in
preparation), thus caution should be taken when compared to available
observational surveys.

\subsection{[Fe/H] and [O/Fe] Profiles of the OHPs}

Previous works have reported the existence of a `gradient' in the
metallicity of stars in the simulated halo systems by considering the
difference between the so-called IHRs and OHRs. In Paper I, we measured
the difference between the median [Fe/H] of the IHRs and OHRs, finding
that simulations were capable of reproducing values similar to those
estimated in the the MW \citep{zolo2009,font2011,mcc2012, tissera2012}.
However, if the analysis is restricted only to the OHRs, one might
expect little or no metallicity gradients, at least if the stars from
the accreted sub-galactic systems were well-mixed. In Paper I, we found that
the median [Fe/H] ([O/Fe]) of our OHPs are correlated (anti-correlated)
with the fractions of stars coming from massive satellites, indicating
that some memory of the history of formation may still remain in this
component. Below we check on this possibility, assuming that their
formation histories are consistent with those predicted by a
hierarchical clustering scenario. 

\begin{figure}
\hspace*{-0.2cm}\resizebox{6.5cm}{!}{\includegraphics{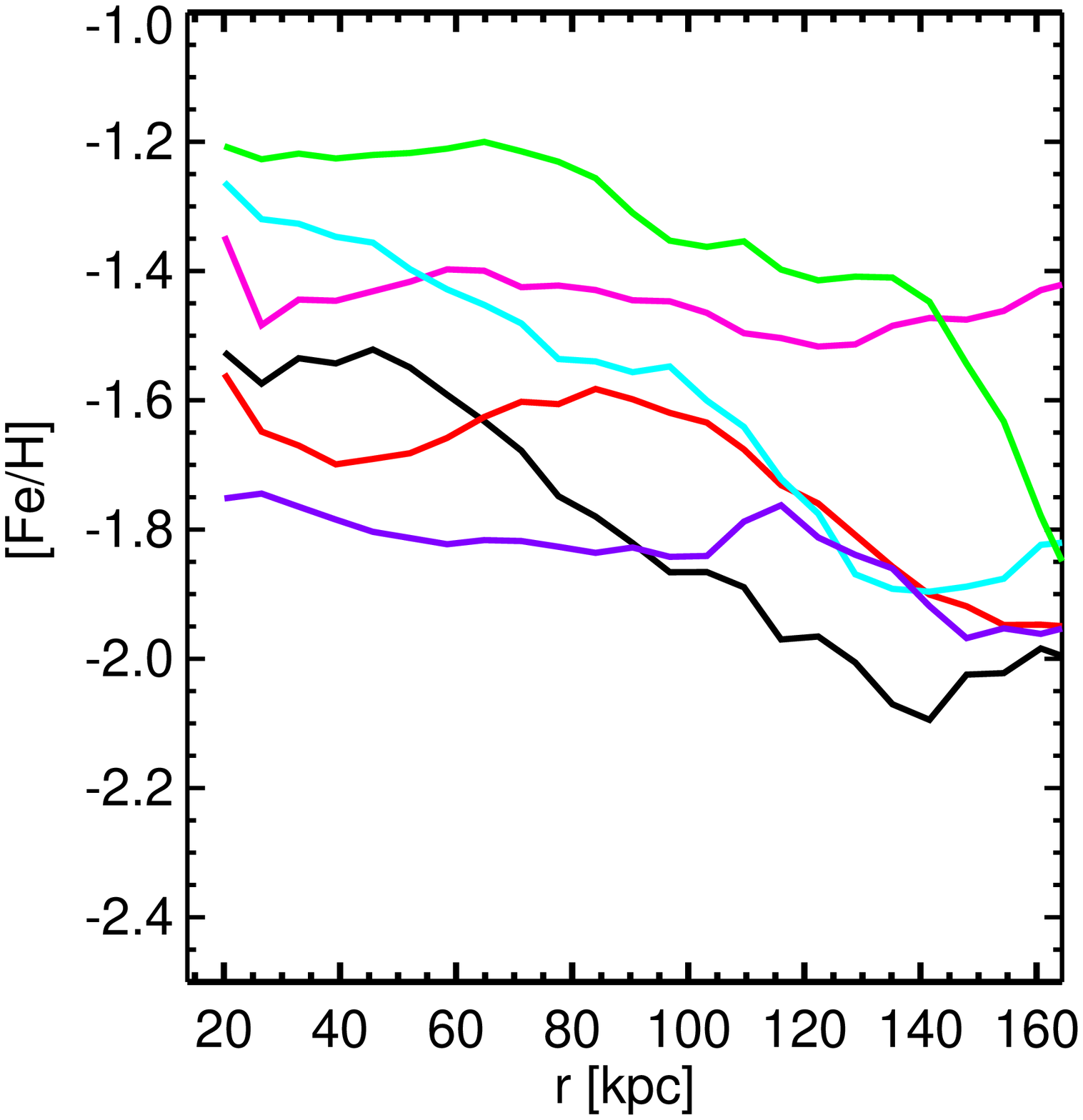}}
\hspace*{-0.2cm}\resizebox{6.5cm}{!}{\includegraphics{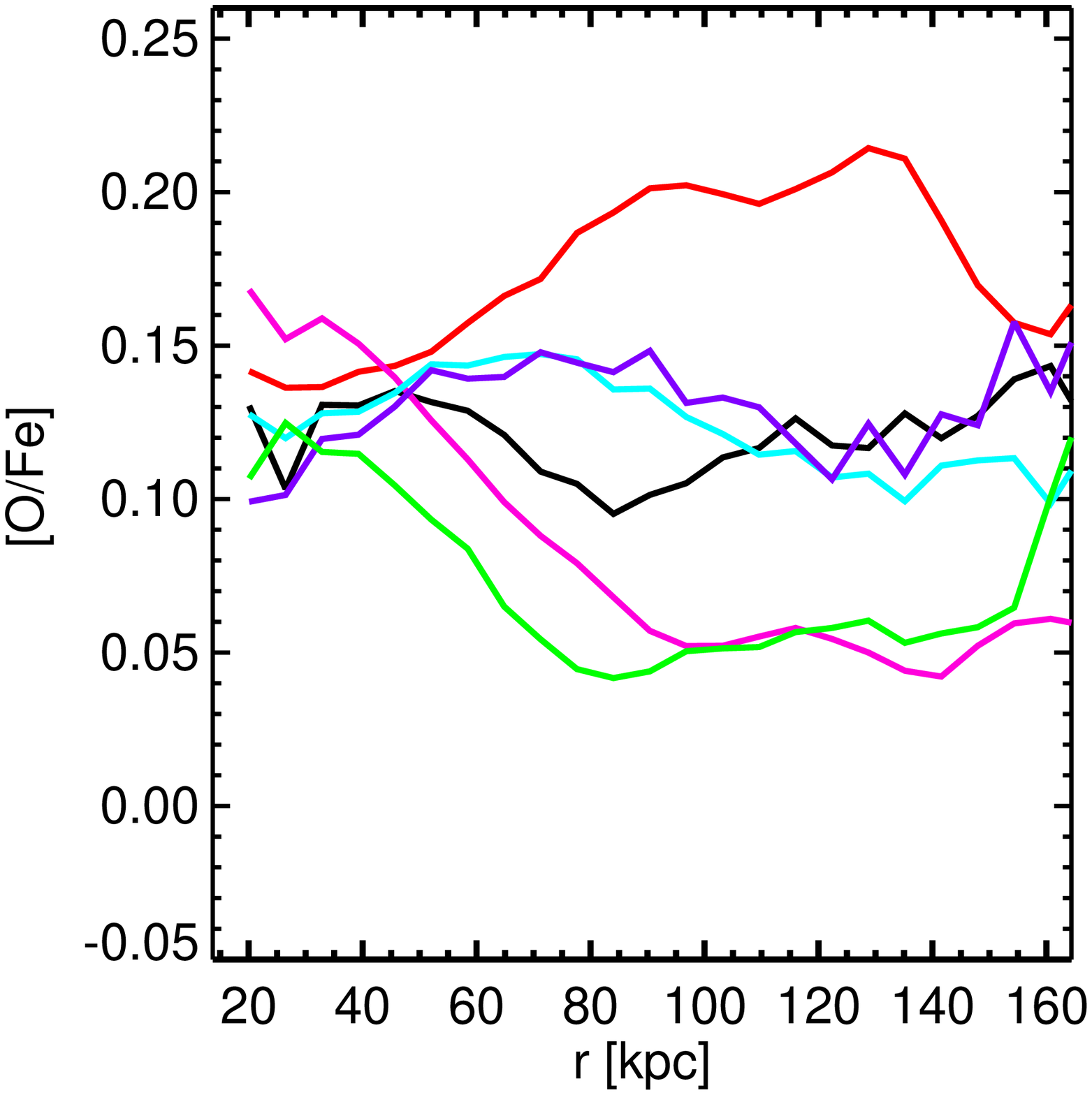}}
\caption{Median abundance profiles of [Fe/H] (upper panel) and [O/Fe] (lower panel) for stars
in the diffuse OHPs of our six simulated galaxies: Aq-A-5 (black), Aq-B-5 (red), Aq-C-5 (magenta),
Aq-D-5 (cyan), Aq-G-5 (green), and Aq-H-5 (violet). See text for
discussion.  }
\label{gradhaloex}
\end{figure}

Fig. ~\ref{gradhaloex} shows the median [Fe/H] and [O/Fe] profiles for
the OHPs in our six simulated halo systems. For [Fe/H], there is
considerable variation in both the level of enrichment and the predicted
slopes with galactocentric distance. We also note that, although this
component has been defined to be smooth, the abundance distributions
might still reveal their complex assembly histories as shown by $Jz$
-$E$ in Paper I \citep[see also][]{helmi2001}. This is particularly
clear from the [$\alpha$/Fe] profiles plotted in the lower panel of
Fig.~\ref{gradhaloex}, where the contributions from the sub-populations
with different levels of $\alpha$-enrichment as a function of
galactocentric radius can be seen. 

\begin{figure}
\resizebox{6cm}{!}{\includegraphics{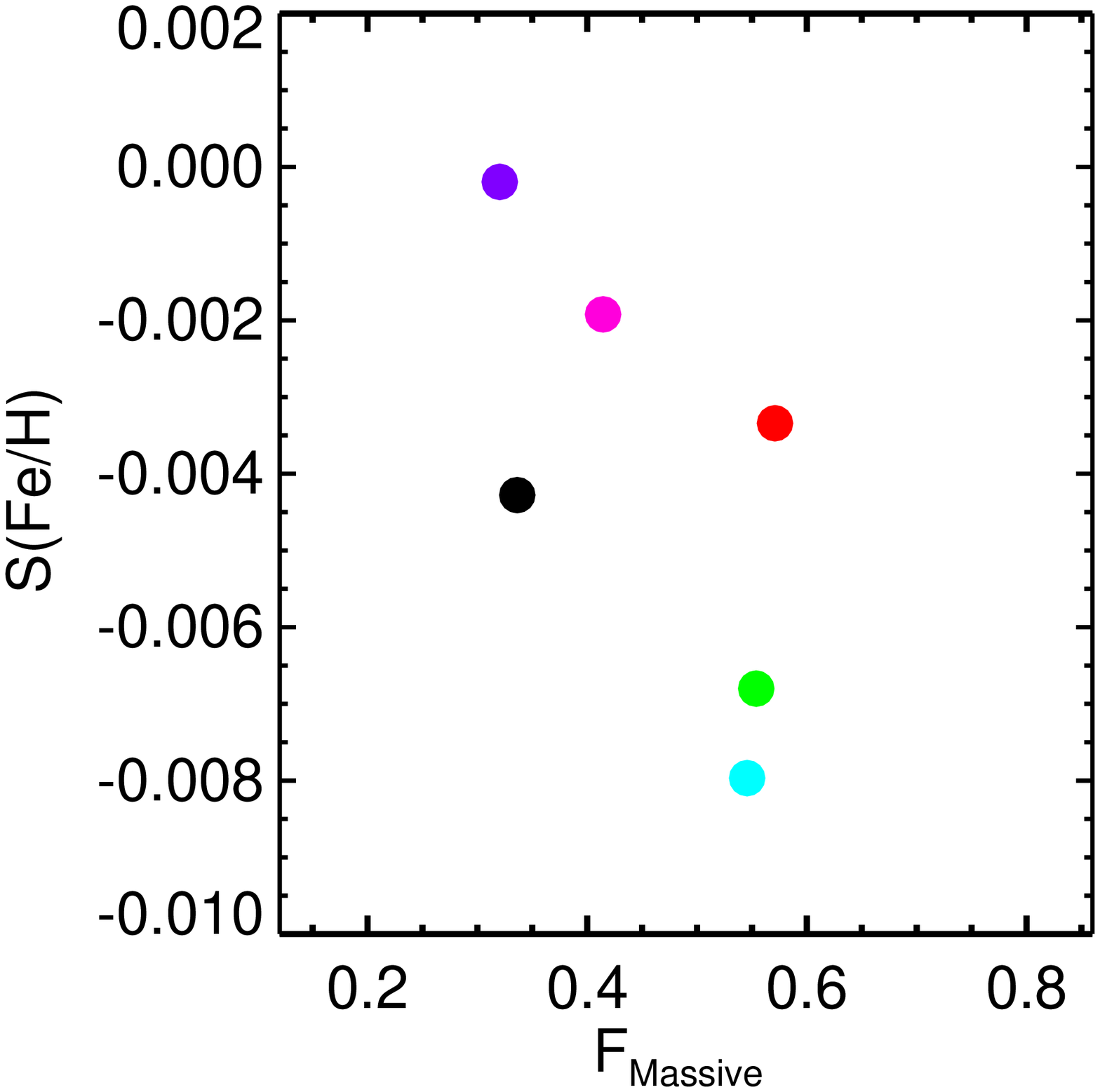}}
\resizebox{6cm}{!}{\includegraphics{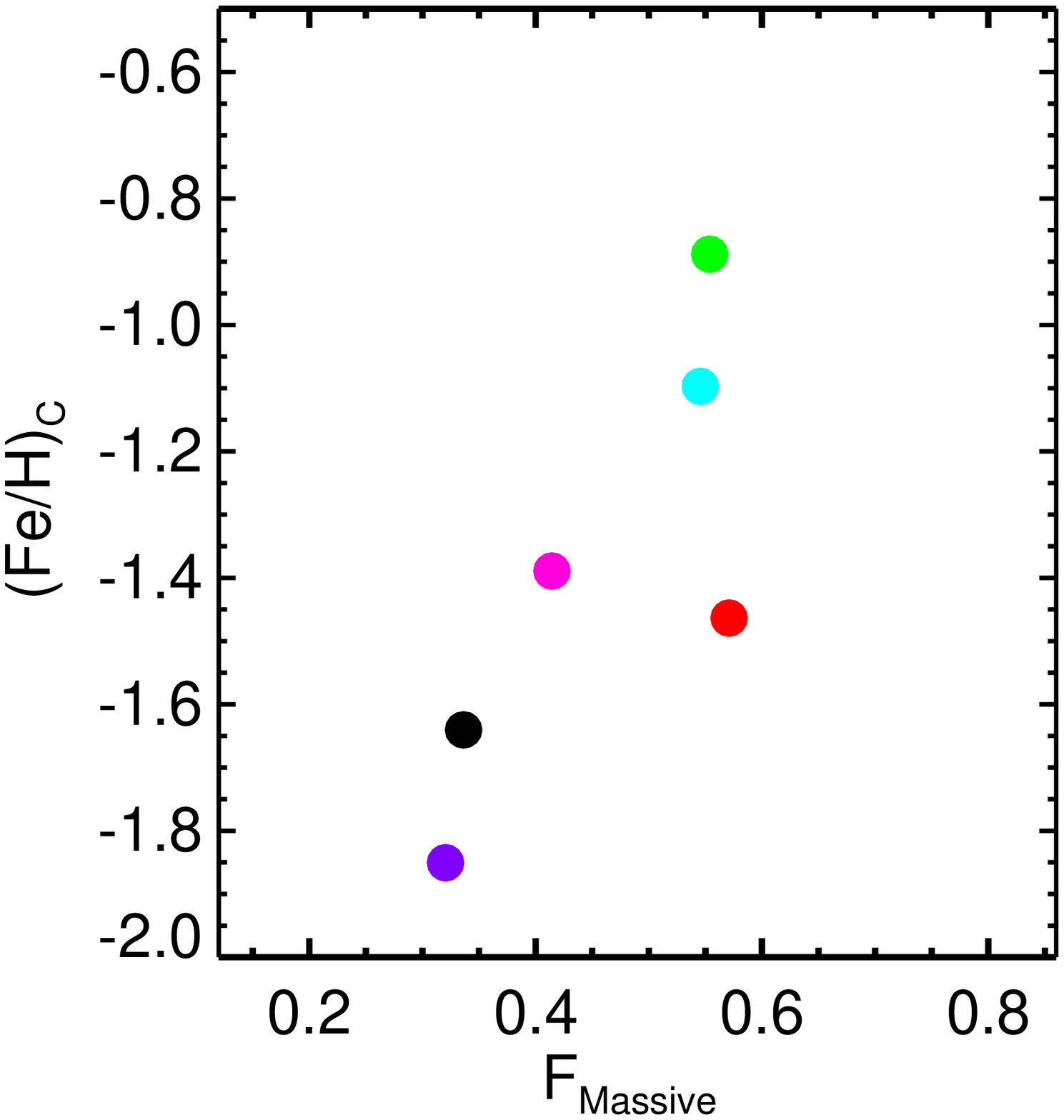}}
\caption{Slopes (upper panel) and zero points (lower panel) of  linear
regressions fit to the [Fe/H] profiles versus the fraction of
stars in the diffuse OHPs formed in satellites more massive than 
$5 \times 10^9M_\odot $h$^{-1}$.  See Fig.~ref{gradhaloex} for colour
coding of the simulated galaxies.}
\label{corre}
\end{figure}

For the [Fe/H] profiles, a linear regression was fit between $\sim 20$
kpc and $\sim 130$ kpc, in order to avoid the central and very outer
parts of the distributions, which might be affected by the IHPs and the
boundaries of the systems, respectively. Fig. ~\ref{corre} shows the
slope and zero points, as a function of the fraction of stars formed in
more massive sub-galactic systems (M $ > 5 \times 10^9$ M$_\odot$). As
expected from the results already discussed in Paper I, the zero point
shows a trend indicating higher metallicities in OHPs formed from larger
contributions of debris stars acquired from relative more-massive
systems. Although the slopes vary from mild to flat, there is a trend
with $F_{\rm massive}$ such that steeper negative slopes are found in
diffuse OHPs formed with significant contributions from relatively
higher-mass systems. Massive sub-galactic systems would tend to survive
farther into the potential wells, carrying relatively more metal-rich
stars and gas available for star formation toward the central regions.
The OHPs received a significant contributions of debris from relatively
lower-mass satellites will tend to be better-mixed, and exhibit more
homogeneous chemical distributions.

\begin{figure*}
\hspace*{-0.2cm}\resizebox{4.5cm}{!}{\includegraphics{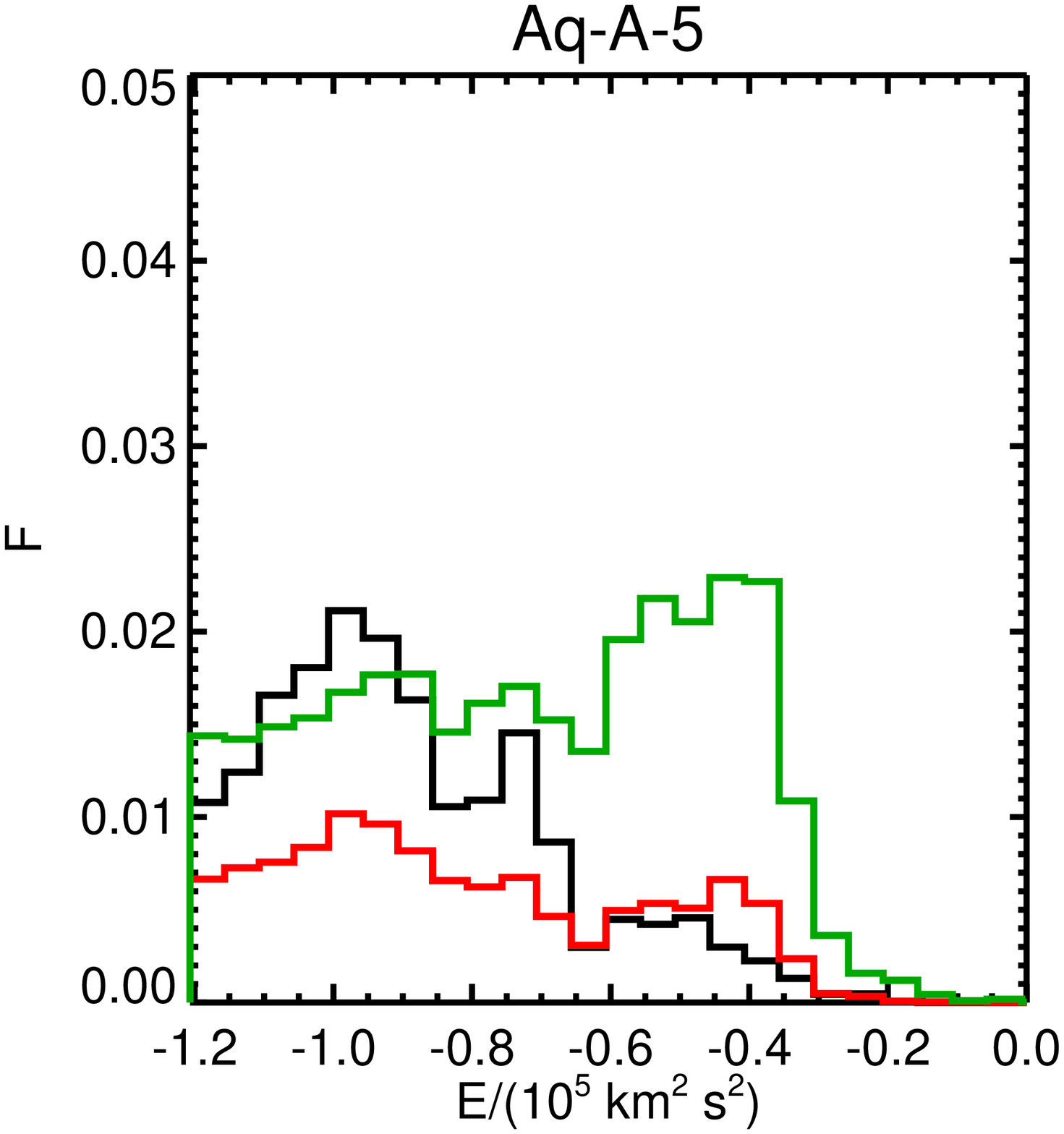}}
\resizebox{4.5cm}{!}{\includegraphics{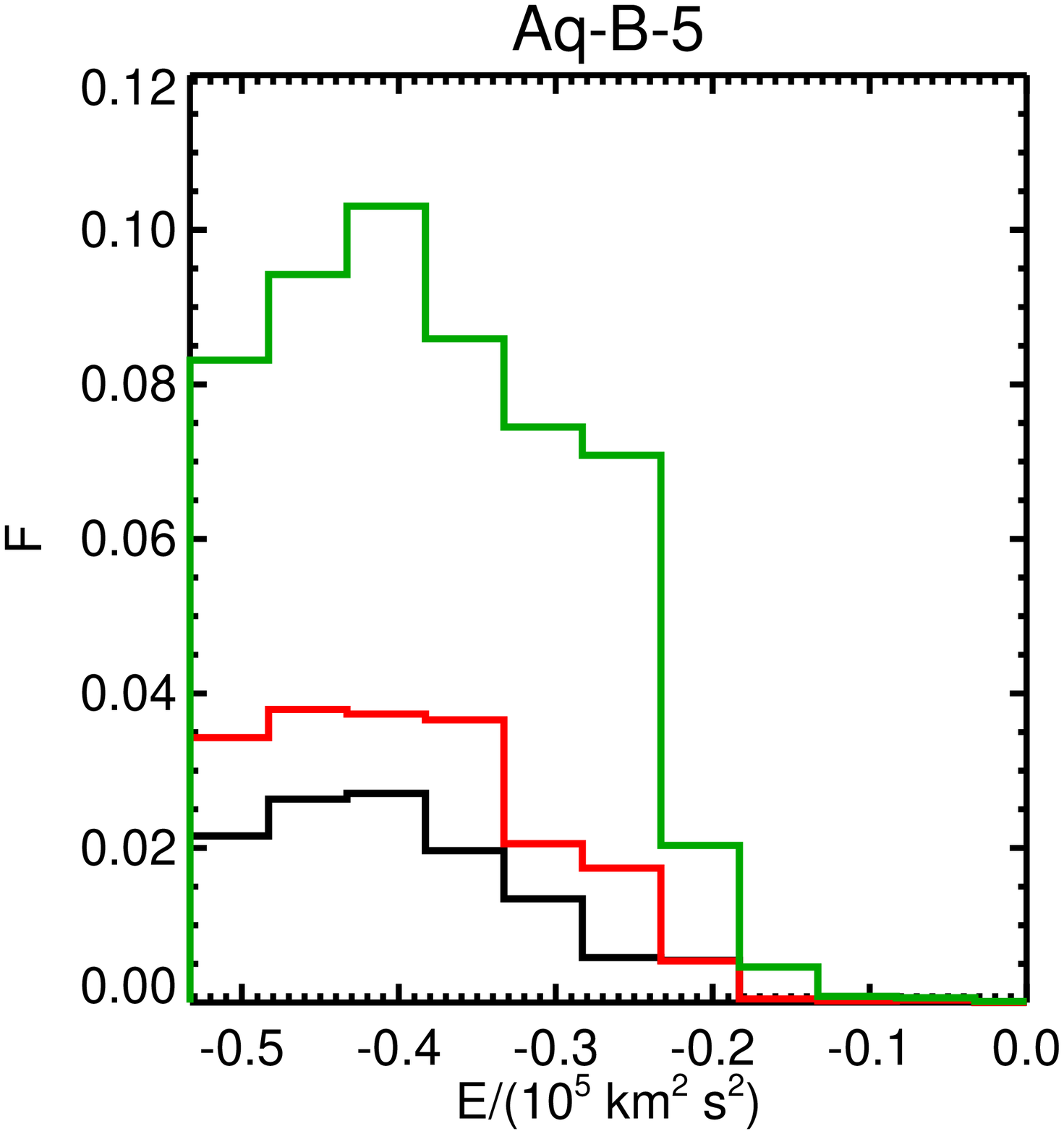}} 
\resizebox{4.5cm}{!}{\includegraphics{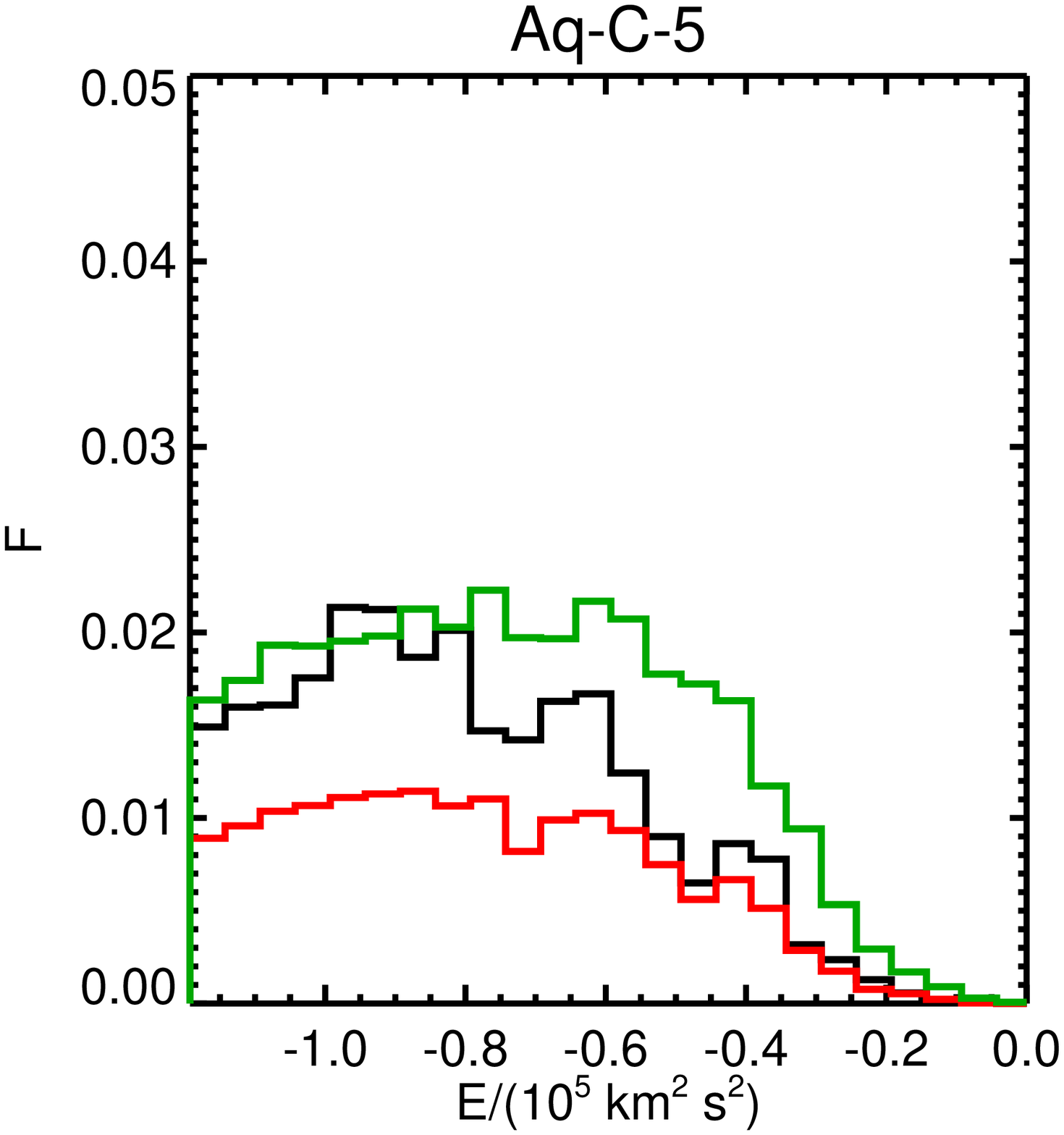}} \\
\hspace*{-0.2cm}\resizebox{4.5cm}{!}{\includegraphics{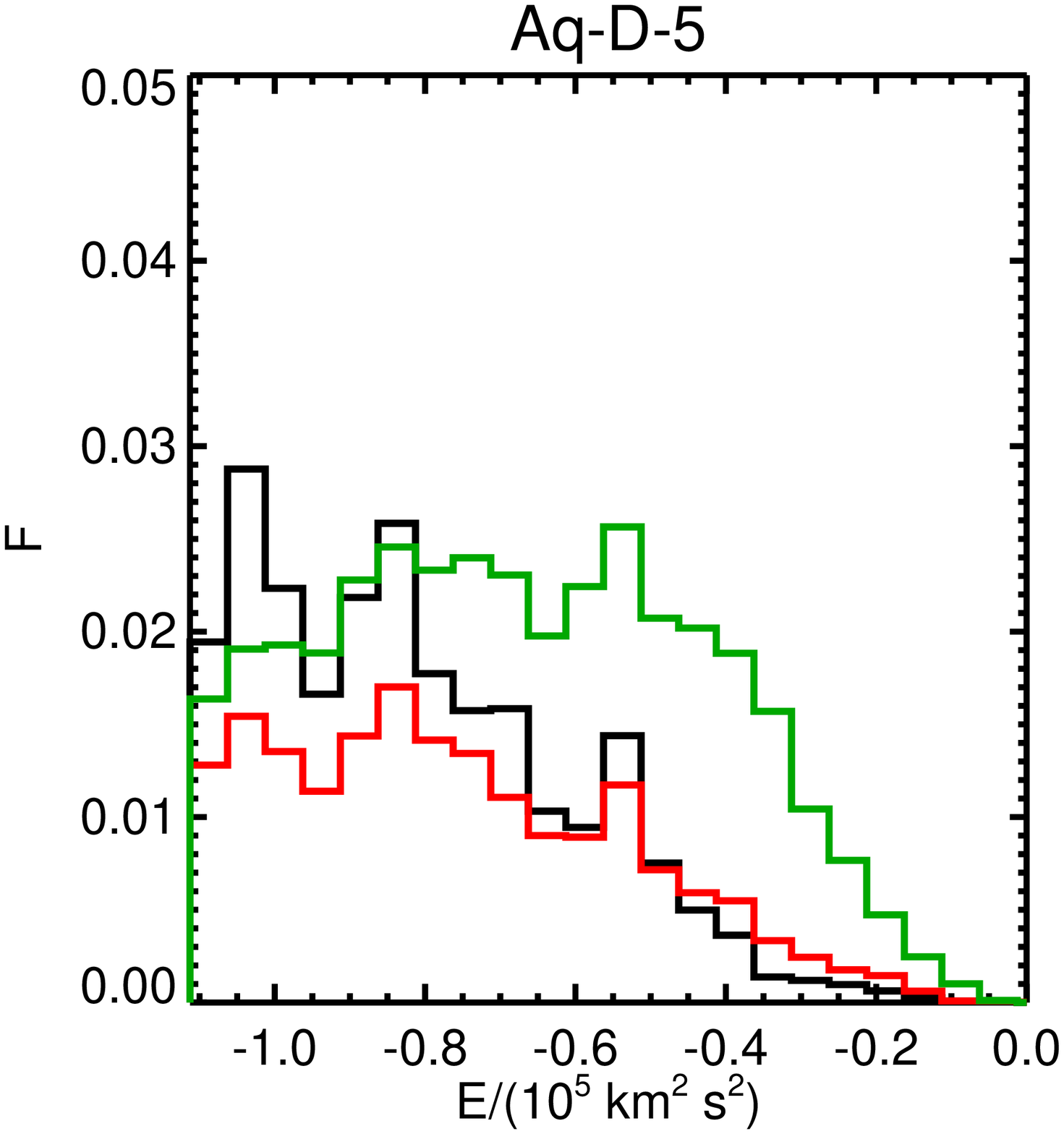}}
\resizebox{4.5cm}{!}{\includegraphics{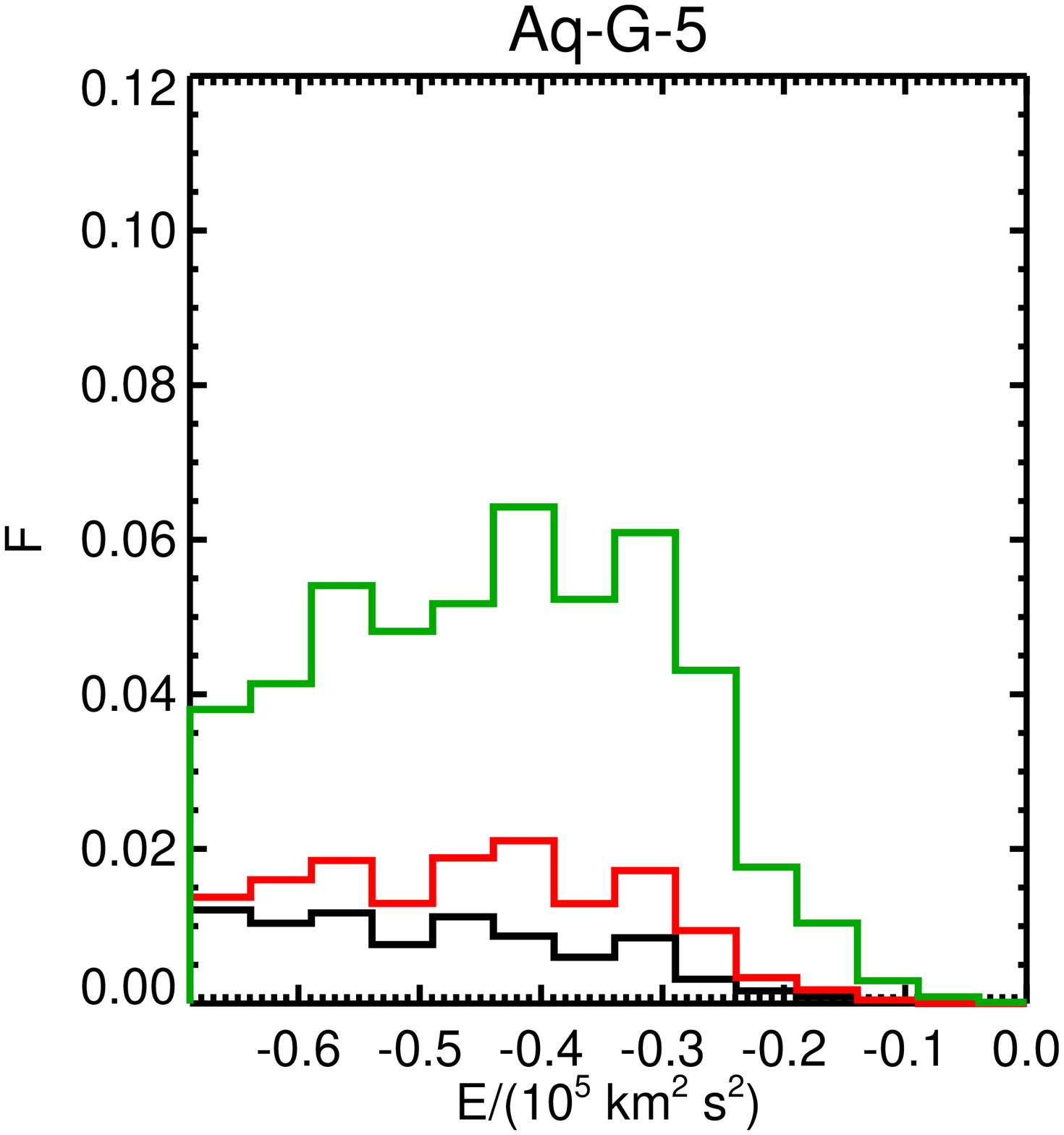}}
\resizebox{4.5cm}{!}{\includegraphics{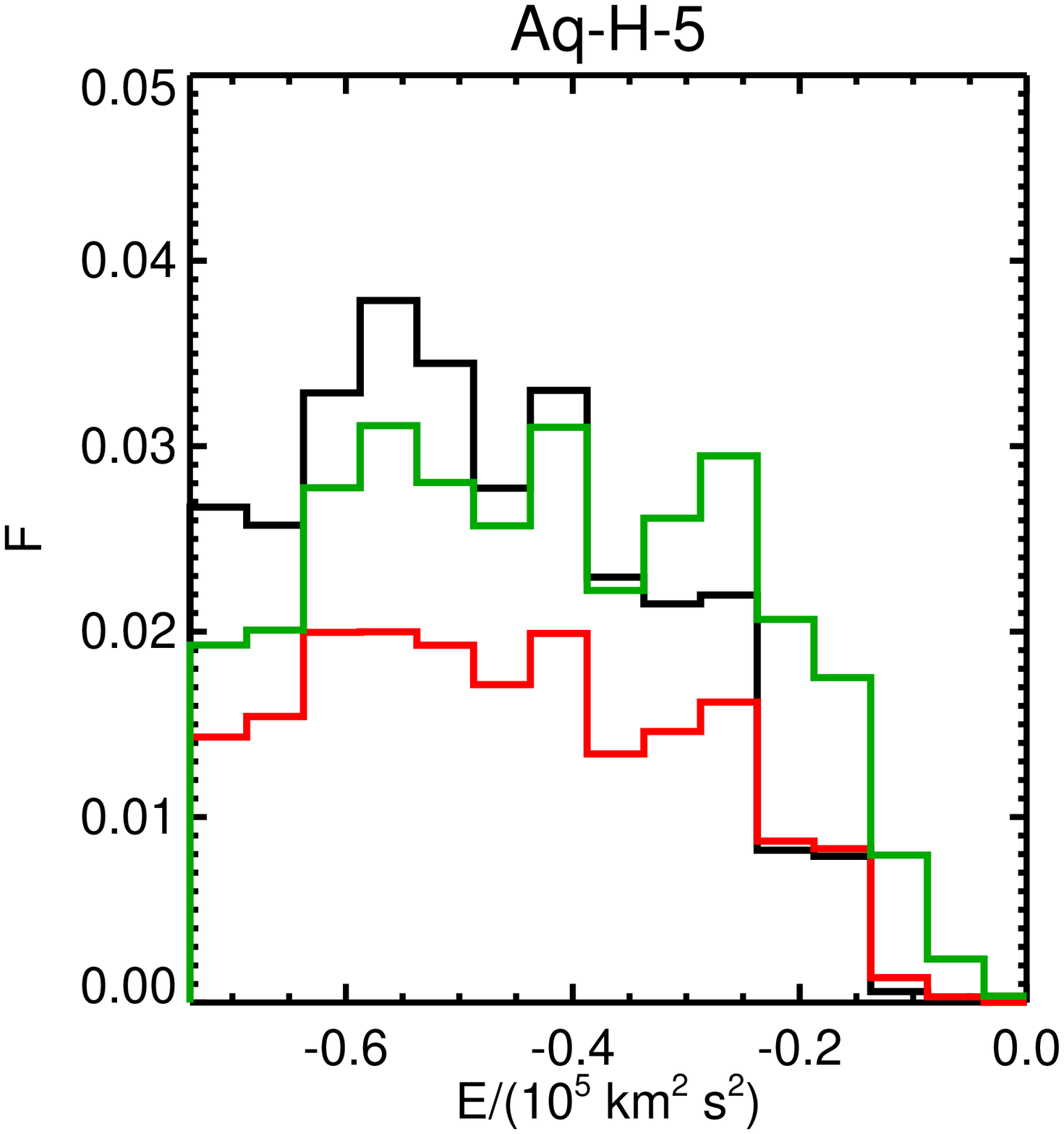}} 
\caption{Distribution of binding energy, $E$, for stars in the diffuse OHPs,  divided according to
their [Fe/H] abundances: [Fe/H]$> -1$ (black lines), $-1.5<$ [Fe/H]$<-1$
(magenta lines) and [Fe/H]$<-1.5$ (green lines).
Aq-A-5 and A-D-5 have steeper metallicity gradients than Aq-C-5 and A-G-5. }
\label{binding}
\end{figure*}

In order to better understand the differences in the abundance
gradients, we also analysed the ages, spatial distributions, and binding
energy distributions of the simulated OHPs. We found that the oldest
stars are less metal-enriched, as expected, and they span a larger range
of total energy and radial distances. Relatively higher-metallicity
stars tend to be more gravitationally bound, and are more centrally
concentrated. The outer portions of the OHPs are dominated by
high-energy stars, which are also the oldest and the lowest metallicity.
To illustrate this, Fig. ~\ref{binding} shows the distribution of
binding energies for stars in the simulated OHPs, segregated according
to [Fe/H]. The two haloes with the flattest metallicity profiles are
those that have similar total energy distributions for stars of
different [Fe/H] abundances. Those with the steeper slopes possess
relatively higher-metallicity stars predominantly at low total energies.
As discussed previously, relatively more-massive satellites contribute
higher-metallicity stars.  Thus, the existence of a gradient in the OHPs is a
  consequence of the properties of the accreted satellites, and how
  they were disrupted within the potential well of the systems. This
  might explain why analytical models grafted onto the purely dynamical
  versions of the Aquarius systems obtained flatter slopes
  \citep{cooper2010, gomez2012}. A comprehensive discussion on the
  metallicity gradients of both IHPs and OHPs is considered in Tissera et al. (in preparation),
  together with  comparisions with recent observations. Accordingly, our results indicate that the
observed metallicity gradient in the diffuse OHR of a galaxy such as the
MW can provide clues about the mass distributions of the satellites that
contributed to its formation.

\section{Conclusions}

We have analysed a suite of six MW mass-sized systems, simulated within
$\Lambda$CDM using a version of {\small P-GADGET3} that includes a
prescription for self-consistent SN feedback. This scheme allows us to
follow the chemo-dynamical evolution of baryons as they are assembled
into galaxies. Our simulated diffuse stellar haloes can be described as
assembled from three  stellar sub-populations: debris, endo-debris and
disc-heated stars. 
Debris stars 
 formed in separate sub-galactic systems located outside the virial radius of
the main progenitor galaxies at their time of formation. These
sub-galactic systems were later accreted by the main progenitor galaxies, donating stars primary to the OHPs, but as well as to
IHPs. The Endo-debris sub-populations formed in gas-rich sub-galactic systems
within the virial radius of the main progenitor galaxies. These
systems were disrupted
farther in the potential well,   contributing stars primarily to the
IHPs, and to a lesser degree, to the OHPs. Disc-heated  stars can
represent a significant percentage of the IHPs.
 These
sub-populations are distinguished from one another as they exhibit, on
average, stars of different chemical abundances, ages, kinematics, and
binding energies. The relative fractions of debris stars and endo-debris
stars provides clues to the mass distribution (and relative number) of
the sub-galactic systems that contribute to form the stellar haloes. These
results have clear implications for observational studies that attempt
to identify different populations in the MW and other galaxies
\citep[e.g.][]{caro2010, caro2012,sheffield2012}.

Summary of the main results:

The simulated IHPs possess three clear stellar sub-populations -- the
debris stars, the endo-debris stars, and the disc-heated stars. Debris
stars are $\alpha$-enhanced relative to iron, low metallicity, and
on average, show no clear signals for rotation. Endo-debris stars exhibit lower
$\alpha$-enhancement than debris ones, on average. Endo-debris stars could
be associated with the stars observed and classified by \citet{sheffield2012}
as accreted stars. In our simulations endo-debris stars exhibit  larger
velocity dispersions, on average,  than the disc-heated stars. Disc-heated stars are
less $\alpha$-enhanced and more metal-rich compared to the endo-debris
stars, with characteristics that depend on their ages. Disc-heated stars
with ages greater then $\sim 10$ Gyr  are dominated by their dispersion,
while those with ages in the range $\sim 8-10$ Gyr have a larger
  rotational velocity component, and exhibit solar to sub-solar [$\alpha$/Fe] ratios. Although
it is feasible, on average, to distinguish these different components by
comparing their chemical abundances and kinematic properties, there
exists considerable overlap between these sub-populations.

The simulated IHPs exhibit metallicity gradients, the strength of which
are determined by the relative fractions of the three sub-populations
that contributed to their formation. While the endo-debris and
disc-heated stars dominate in the central parts of IHRs (within $\approx
20$ kpc), the debris stars dominate in the outer parts of the IHRs.
Hence, the relative contributions of the different sub-populations can
conspire to produce steeper metallicity gradients than those exhibit by
each sub-populations separately.
  
The simulated OHPs are primarily comprised of debris stars that formed
in accreted sub-galactic systems. The relatively less-massive systems are
easily disrupted, while the relatively more-massive ones are able to
survive farther into the global potential well of the system before
being disrupted. As a consequence, although stars of the OHPs are
generally of very low metallicity, the relatively more metal-rich ones
tend to be more centrally concentrated. Some of the OHPs show
  signals of weak prograte rotation ($\approx 15\%$  V$_{\rm LSR}$).The relative distribution of
debris stars from these different accreted sub-galactic systems can result
in a genuine metallicity gradient within the OHRs, which although weak,
can be used to constrain the assembly history of the OHPs. The
[$\alpha$/Fe] ratios exhibit a heterogeneous distribution from simulated
galaxy-to-galaxy, reflecting the mixture of contributions from accreted
galactic systems with different star-formation histories. A future
comparison between the frequency of low-[$\alpha$/Fe] stars in the
haloes can provide clues on the history of formation and the masses of
the accreted systems, as well as on the validity of current
chemical-enrichment schemes.

In this paper, we have studied the IHPs and OHPs as separate
sub-populations in relation to their assembly histories in a
cosmological context. The properties of the combined IHPs and OHPs will
be discussed in a forthcoming paper, which considers comparisons of the
nature of the observed stellar populations in the MW, M31, and and other
nearby galaxies with predictions from the simulations.

\section*{Acknowledgments}

We thank Simon D.M. White for useful comments and suggestions.
This work was partially funded by L'Oreal-Unesco-Conicet Award for
Women in Science, PICT Max Planck 245 (2006) of the Ministry of Science
and Technology (Argentina), and the Cosmocomp and Lacegal Networks of
FP7 Programme of the European Community.  TCB acknowledges partial
funding from grants PHY 02-16783 and PHY 08-22648:  Physics Frontier
Center / Joint Institute for Nuclear Astrophysics (JINA), awarded by the
U.S. National Science Foundation.


\end{document}